\documentclass[letterpaper,twocolumn,10pt]{article}
\usepackage{usenix}

\usepackage{tikz}
\usepackage{amsmath}
\usepackage{amssymb}
\usepackage{multicol}
\usepackage{multirow}

\usepackage{filecontents}
\begin{document}

\date{}

\title{\Large \bf SoK: Can Synthetic Images Replace Real Data?\\
A Survey of Utility and Privacy of Synthetic Image Generation}

\author{
{\rm Yunsung Chung \qquad Yunbei Zhang \qquad Nassir Marrouche \qquad Jihun Hamm}\\
Tulane University
} 

\maketitle

\begin{abstract}
Advances in generative models have transformed the field of synthetic image generation for privacy-preserving data synthesis (PPDS). However, the field lacks a comprehensive survey and comparison of synthetic image generation methods across diverse settings. 
In particular, when we generate synthetic images for the purpose of training a classifier, there is a pipeline of generation-sampling-classification which takes private training as input and outputs the final classifier of interest. 
In this survey, we systematically categorize existing image synthesis methods, privacy attacks, and mitigations along this generation-sampling-classification pipeline. 
To empirically compare diverse synthesis approaches, we provide a benchmark with representative generative methods and use model-agnostic membership inference attacks (MIAs) as a measure of privacy risk. 
Through this study, we seek to answer critical questions in PPDS: Can synthetic data effectively replace real data? Which release strategy balances utility and privacy? Do mitigations improve the utility-privacy tradeoff? Which generative models perform best across different scenarios? 
With a systematic evaluation of diverse methods, our study provides actionable insights into the utility-privacy tradeoffs of synthetic data generation methods and guides the decision on optimal data releasing strategies for real-world applications.
\end{abstract}

\section{Introduction}

Can synthetic data replace real data? This question has become increasingly relevant as the demand for privacy-preserving techniques grows across various domains such as healthcare, finance, and social media~\cite{abouelmehdi2018big, beigi2018privacy, khalid2023privacy, lata2023deep, williamson2024balancing}. In these fields, Privacy-Preserving Data Sharing (PPDS) offers a promising approach to protecting privacy while retaining analytical value. For example, in healthcare where patient data must be safeguarded, PPDS frameworks can enable machine learning model training on synthetic data without exposing real patient information. Similarly, in finance, synthetic transaction data allows the development of fraud detection systems while preserving customer confidentiality. These highlight the potential of PPDS to protect privacy while maintaining the utility. However, achieving an optimal balance of synthetic data between utility (such as data fidelity and classification accuracy) and privacy (such as Differential Private (DP)~\cite{dwork2014algorithmic} and Membership Privacy~\cite{izzo2022provable}) remains a significant challenge~\cite{stadler2022synthetic}. 

This paper seeks to answer fundamental questions about the role of synthetic data in PPDS: (1) Can synthetic data effectively replace real data? (2) Which is safer to share—synthetic images or classifiers trained on synthetic data? (3) Do mitigation strategies improve utility-privacy tradeoffs? (4) Which generative models perform best across different data-sharing scenarios? By addressing these questions, we aim to provide actionable insights into the strengths, limitations, and practical applicability of synthetic data for real-world privacy-preserving pipelines.

Generative models such as GANs~\cite{goodfellow2014generative, mirza2014conditional, radford2015unsupervised, arjovsky2017wassersteingan, karras2017progressive, brock2018large, zhang2019self, karras2020training}, VAEs~\cite{kingma2013auto, chen2016variational, van2017neural, razavi2019generating, child2020very}, and diffusion models~\cite{ho2020denoising, song2020denoising, dhariwal2021diffusion, ho2022cascaded, ramesh2022hierarchical, rombach2022high} have significantly advanced synthetic data generation. 
Subsequently, privacy-enhancing techniques such as DP~\cite{dwork2014algorithmic} and DP-SGD~\cite{abadi2016deep} have been integrated and tested with generative models. These methods aim to ensure robust privacy guarantees by controlling the contribution of individual samples during training. Studies combining DP with diffusion models~\cite{dockhorn2022differentially, ghalebikesabi2023differentially, lyu2023differentially, wang2024dp} have shown potential for generating privacy-preserving synthetic images. Similarly, adversarial training and mitigation strategies~\cite{luo2024privacy} have emerged to address the utility-privacy tradeoffs of generative models. However, the effectiveness of these methods in mitigating privacy risk across different release strategies and data-sharing settings remains unclear.

While there are a number of surveys and benchmarks on the utility-privacy tradeoff of tabular data~\cite{stadler2022synthetic, sarmin2024synthetic}, studies on synthetic image data are rare. A recent Systematization of Knowledge (SoK) on image generation~\cite{hu2024sok} highlighted the challenges in producing high-quality privacy-preserving synthetic images under practical privacy budgets, which emphasizes the need for further study in systematically evaluating utility-privacy tradeoffs. Furthermore, direct comparison across diverse generative models such as GANs, diffusion models, and data condensation~\cite{wang2018dataset, zhao2020dataset, zhao2023dataset} within a unified framework is rarely found in the literature due to the difficulty of consistent evaluation. The few existing empirical studies are also inconclusive on the current success of synthetic image generation. For instance, while a prior work~\cite{stadler2022synthetic} reports significant utility loss when applying privacy-preserving techniques, \cite{sarmin2024synthetic} reported improved tradeoffs after addressing experimental flaws of the former. These observations underscore a critical need for a comprehensive systematization that not only surveys the landscape but also empirically benchmarks diverse approaches to provide clear and actionable insights.

\begin{figure*}
    \centering
    \includegraphics[width=.90\linewidth]{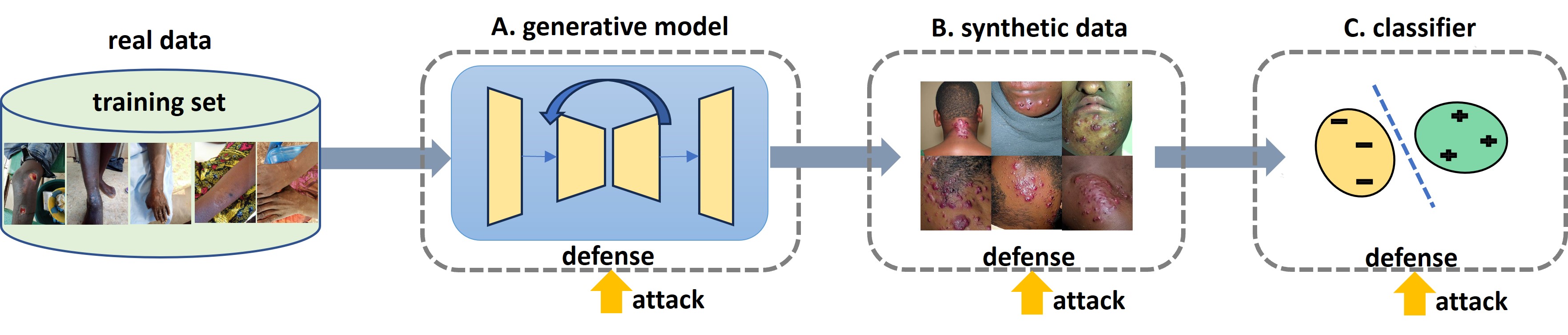}
    \caption{The generation-classification pipeline for sharing synthetic images. Privacy data can be shared as (A) a trained generative model, (B) synthetized images, or (C) a classifier trained on synthetic data. Each type of shared information involves distinct privacy attacks and mitigation strategies with corresponding differences in utility-privacy tradeoffs.
    }
    \label{fig:schematic}
\end{figure*}

To address these gaps, this paper presents a SoK on the utility-privacy tradeoffs in privacy-preserving synthetic image generation. Our work goes beyond existing surveys by:
\begin{enumerate}
    \item \textbf{Developing and Applying a Unifying Framework}: We introduce a data-sharing pipeline (Figure~\ref{fig:schematic}, detailed in Sec.\ref{sec:problem setting}) that serves as a framework to systematically categorize and analyze methods, privacy attacks, and mitigation strategies at distinct points of information release: (1) the generative models; (2) the synthetic images produced, or (3) classifiers trained on synthetic data. This structured approach (detailed in Secs.~\ref{sec:generative models}, \ref{sec:synthetic images}, and \ref{sec:classifiers}) allows for a coherent understanding of he varying risks and benefits.
    \item \textbf{Conducting a Broad and Unified Empirical Evaluation}: We conduct a large-scale empirical study (Sec.~\ref{sec:benchmark}) assessing a wide range of representative approaches selected to cover diverse generative architectures (VAE, GANs, Diffusion models, data condensation), training paradigms (from scratch vs. fine-tuning), and privacy considerations (non-mitigated, DP-enhanced, and other mitigation strategies). This evaluation, which represents a significant computational undertaking (detailed runtime analyses are provided in Appendix~\ref{tab:chexpert_running_times} to contextualize the cost), is performed under consistent conditions using established metrics for utility (classification performance) and privacy (Membership Inference Attacks) to allow for direct and fair comparisons—a critical element often missing in prior works.
    \item \textbf{Critically Evaluating and Contextualizing Knowledge}: Through our benchmark, we provide novel insights into the practical performance, strengths, and limitations of these varied approaches across different datasets (CelebA, Fitzpatrick17k, and CheXpert) and data-sharing scenarios. This allows us to contextualize theoretical claims and evaluate the real-world efficacy of different utility-privacy balancing strategies.
    \item \textbf{Identifying Actionable Insights and Guiding Future Research}: Based on our systematic findings, we derive actionable insights for practitioners and identify open challenges and promising directions for future research in this domain.
\end{enumerate}
Our literature search strategy to inform the selection of methods for this systematization involved employing relevant keywords (e.g., "membership inference attack," "generative models," "differential privacy," "synthetic data generation") on Google Scholar, prioritizing influential and recent works to ensure diverse coverage across methods and pipeline stages. This comprehensive approach, combining a structured framework with extensive empirical evidence, aims to bridge the gap between theory and practice in privacy-preserving synthetic image generation.


\paragraph{Summary of Findings.}
Our comprehensive evaluation of representative methods across diverse datasets (\emph{e.g.}, CelebA~\cite{liu2015faceattributes}, Fitzpatrick17k~\cite{groh2021evaluating}, and CheXpert~\cite{irvin2019chexpert}) yields several critical insights into privacy-preserving synthetic data sharing (detailed in Sec.~\ref{sec:benchmark}):
\begin{enumerate}
    \item The optimal data-sharing strategy (releasing synthetic images vs. classifiers trained upon them) is \textbf{highly dataset-dependent}: while classifiers often provide better utility-privacy tradeoffs for visually diverse datasets (\emph{e.g.}, CelebA, Fitzpatrick17k), direct image sharing can be less risky for certain methods on more homogeneous data (\emph{e.g.}, CheXpert).
    \item Classifiers trained on high-quality synthetic data, particularly from \textbf{advanced diffusion models}, can achieve \textbf{superior utility-privacy balances} compared to classifiers trained directly on real data with DP-SGD, sometimes even offering better performance for underrepresented subgroups.
    \item Effective privacy can be achieved not only through \textbf{explicit mitigation techniques} (\emph{e.g.}, SMP-LoRA showing strong utility-privacy balance) but also via \textbf{implicit strategies}, such as careful hyperparameter tuning or specific generative processes (\emph{e.g.}, multi-instance synthesis with TI/LoRA surprisingly reducing individual sample memorization).
    \item The choice of \textbf{generative model architecture is crucial}, with diffusion-based methods generally outperforming the tested GAN and VAE models in achieving favorable utility-privacy tradeoffs and image fidelity, though the relationship between utility and privacy is not always a simple inverse.
\end{enumerate}
Through these findings, we provide actionable insights for designing privacy-preserving strategies tailored to specific dataset characteristics, utility requirements, privacy goals, and fairness considerations within PPDS.

The rest of the paper is organized as follows. We first review related work in Sec. \ref{sec:related work} to identify existing research gaps in PPDS. In Sec. \ref{sec:problem setting}, we describe the synthetic image generation and classification setting with introducing the generation-classification pipeline. Following the pipeline, Secs. \ref{sec:generative models}, \ref{sec:synthetic images}, and \ref{sec:classifiers} provide a comprehensive survey for each stage. Our benchmarking study in Sec. \ref{sec:benchmark} evaluates privacy-utility tradeoffs and presents our findings. We conclude in Sec. \ref{sec:conclusion} with insights from the experiments.

\section{Related Work} \label{sec:related work}

\paragraph{PPDS for Tabular Data.}
Several works have investigated PPDS in tabular data settings. Stadler et al.~\cite{stadler2022synthetic} examined synthetic data release methods, observing frequent challenges in balancing utility and privacy. Sarmin et al.~\cite{sarmin2024synthetic} revisited this issue to address limitations in experimental design and report more favorable utility-privacy outcomes with adjustments. Qian et al.~\cite{qian2023synthcity} introduced a resource for generating and evaluating synthetic tabular data. Lautrup et al.~\cite{lautrup2024syntheval} proposed a framework emphasizing comprehensive privacy and utility evaluation for tabular datasets. Tao et al.~\cite{tao2021benchmarking} compared DP synthetic data methods with measuring utility with distributional and machine learning metrics.

While these contributions have advanced PPDS in tabular data, the adaptability to synthetic image generation remains limited. We address this gap by providing a focused analysis of PPDS within the context of synthetic image data generation.

\paragraph{PPDS for Image Data.}
Several reviews explore PPDS specifically within generative modeling. Hu et al.\cite{hu2024sok} provided a detailed review of differentially private data generation techniques and included a small-scale study of six models on MNIST and FMNIST images. Their focus is exclusively on DP and not on MIAs. Troung et al.\cite{truong2024attacks} surveyed attack and defense strategies for diffusion models including MIAs, adversarial attacks, and backdoor attacks. However, this analysis did not extend to any comparative experiments across models.
In contrast, we use practical and real-world datasets including Fitzpatrick and CelebA in limited data sizes. Additionally, we cover an extensive range of generative models, including GANs, VAEs, diffusion models, and data condensation methods. This broader scope enables a more comprehensive analysis of PPDS relevant to real-world applications.

\paragraph{MIA and Mitigation Methods in Generative Modeling.}
Membership inference attacks and their mitigations have been systematically studied within traditional machine learning contexts. Salem et al.\cite{salem2018ml} developed ML-Leaks to analyze MIAs in classifier models, assessing adversarial setups, classifier types, datasets, and defense techniques such as dropout and model stacking. Hu et al.\cite{hu2022membership} reviewed MIAs in various machine learning models, with some emphasis on generative models but without detailed experimental comparison. Niu et al.~\cite{niu2023sok} introduced MIBench, an extensive MIA evaluation framework covering numerous configurations across datasets and model architectures.

Recent work has proposed MIAs tailored to generative models, including GANs and diffusion models~\cite{pang2023black, pang2023white, van2023membership, zhang2024generated}. However, These studies tend to focus on limited comparisons with non-defended models, which underscores the need for a broader analysis that incorporates attacks and defenses within synthetic image generation.

\section{Problem Setting} \label{sec:problem setting}

We describe the setting of synthetic image generation and classification, and privacy attacks on the generated images and classifiers. 

\subsection{Image Generation-Classification Pipeline.}
In the privacy-preserving data generation scenario illustrated in Figure~\ref{fig:schematic}, a generative model $g$ is trained on a private dataset $D_{tr}$ sampled from an underlying distribution $p_R$. This generative model $g$ may be either directly trained on $D_{tr}$ or pre-trained on an independent dataset and fine-tuned with $D_{tr}$. The output distribution of this model is denoted as $p_G$ after training.

The generative model $g$ produces a synthetic dataset $D_{syn}$, sampled from $p_G$. This synthetic data serves as a privacy-preserving alternative to the original $D_{tr}$ and can be shared publicly or used to train downstream models. Following this, a classifier $f$ is trained on $D_{syn}$, with the option of initial pre-training on a dataset independent of both $D_{tr}$ and $D_{syn}$. This pipeline of generating and sharing synthetic data and classifiers is common in scenarios where the protection of sensitive training data $D_{tr}$ is significant.

The choice of which artifact to release fundamentally alters the attack surfaces and the applicable defense mechanisms. These artifacts can be the generative model $g$ (Stage A), the synthetic dataset $D_{syn}$ (Stage B), or the classifier $f$ (Stage C). We explore a systematization of these choices in Secs \ref{sec:generative models}, \ref{sec:synthetic images}, and \ref{sec:classifiers}.
\subsection{Privacy Attack on the Pipeline.}
MIAs aim to determine if a specific data sample $x$ was part of the training set $D_{tr}$ used for training a model. In MIAs, an adversary attempts to classify either members $x\in D_m = D_{tr}$ or non-members $x \in D_{nm} = D \setminus D_{tr}$. MIAs are widely used in privacy research because they provide a direct way to assess privacy leakage in machine learning models~\cite{shokri2017membership, hayes2017logan, hilprecht2019monte, carlini2022membership}. MIAs are often regarded as a practical standard for testing the privacy of machine learning models
due to their relative simplicity in implementation compared to the other types of attacks. They provide a binary metric indicating potential training data inclusion.

One of the key advantages of MIAs is they can be applied at any stage in the generation-classification pipeline (Figure~\ref{fig:schematic}). This flexibility enables MIAs to assess privacy risks in generative models, synthetic images, or classifiers. For instance, when synthetic data is shared, adversaries utilize synthetic data distribution $p_S$ or its similarity with query data $x$ to infer membership information. Furthermore, classifiers trained on $D_{syn}$ may have traces of $D_{tr}$, which leads adversaries to utilize classifier outputs for membership information.

The specific potency of these MIAs varies depending on the pipeline stage where information is exposed, as detailed in our systematization of attacks in Secs \ref{sec:generative-model-attacks}, \ref{sec:image-attacks}, and \ref{sec:classifier-attacks}.

\paragraph{Auxiliary Data Assumptions.}
In privacy studies, the level of risk posed by an adversary often depends on the types of auxiliary data they have access to. Auxiliary data $D_{aux}$ refers to additional information the adversary might use to infer private information. We highlight three main categories:\\
\textbf{No Auxiliary Data}: The adversary has only access to the synthetic data $D_{syn}$ or a classifier $f$ trained on $D_{syn}$~\cite{hayes2017logan, hilprecht2019monte, chen2020gan, carlini2022membership}.\\
\textbf{External Auxiliary Data}: The adversary has access to auxiliary data $D_{aux}$ sampled from similar but non-overlapping with $D_{tr}$~\cite{shokri2017membership, van2023membership, zhang2024generated, pang2023black}.\\
\textbf{Subset Auxiliary Data}: The adversary's $D_{aux}$ includes members and non-members from $D_{tr}$, which represents the strongest privacy challenge by offering direct exposure to $D_{tr}$~\cite{carlini2022membership, wu2022membership, pang2023black}.\\
In this survey, we include approaches that encompass all three assumptions. In Sec.~\ref{sec:benchmark}, we use the last assumption which is the most challenging setting for privacy preservation.


\subsection{Privacy preservation on the pipeline}

Privacy-preserving techniques, or mitigation methods, aim to reduce an adversary’s ability to infer membership of the training data by reducing privacy leakage of the shared information. In this work, in particular, mitigation methods can be applied to reduce the MIA success at three different stages of the generation-classification pipeline as illustrated in Fig.~\ref{fig:schematic} -- the generative model $g$, the synthetic dataset $D_{syn}$, or the classifier $f$. Although it may be unnecessary, multiple mitigation methods can also be applied at multiple stages. 

Our survey systematically categorizes these mitigation techniques based on their point of application within the pipeline (Secs. \ref{sec:generative-model-mitigation}, \ref{sec:image-mitigation}, and \ref{sec:classifier-mitigation}) and analyzes their underlying principles and effectiveness against the corresponding attacks.

\begin{table*}[ht]
\scriptsize
\caption{Summary of Methods: This table categorizes and lists the Models, Attacks, and Mitigation methods for each stage of the pipeline as illustrated in Figure \ref{fig:schematic}. Note that any model can be utilized for training the classifier.
}
\label{table:summary}
\centering
\begin{tabular}{p{1.1cm}|p{2.8cm}|p{3.3cm}||p{1.5cm}|p{1.6cm}||p{2.5cm}|p{1.7cm}}
\hline
\rule{0pt}{2.6ex}\multirow{2}{*}{\textbf{}} & \multicolumn{2}{|c||}{\textbf{Release Generative Model (A)}} & \multicolumn{2}{c||}{\textbf{Release Synthetic Image (B)}} & \multicolumn{2}{|c}{\textbf{Release Classifier (C)}} \\
\cline{2-7}
\rule{0pt}{2.6ex}& \textbf{Category} &  \textbf{Reference} &  \textbf{Category} &  \textbf{Reference} &  \textbf{Category} & \textbf{Reference} \\

\hline

\rule{0pt}{2.6ex}\multirow{4}{*}{\textbf{Model}} & GAN/VAE & \cite{goodfellow2014generative,mirza2014conditional, radford2015unsupervised, karras2017progressive, zhang2019self, brock2018large, salimans2018improving, chen2016variational, van2017neural, razavi2019generating, genevay2018learning, patrini2020sinkhorn, deja2020end} & & & \multirow{4}{*}{-} & \multirow{4}{*}{-} \\
\cline{2-3}
\rule{0pt}{2.6ex}& Finetuned GAN/VAE & \cite{mo2020freeze, karras2020training} & Data \newline Condensation & \cite{wang2018dataset, zhao2020dataset, nguyen2020dataset, zhao2021dataset, nguyen2021dataset, cazenavette2022dataset, zhao2023dataset} & & \\
\cline{2-3}
\rule{0pt}{2.6ex}& Diffusion Models & \cite{ho2020denoising,song2020denoising,song2019generative, song2020improved, song2020score, dhariwal2021diffusion, batzolis2021conditional,rombach2022high}  & & & &\\
\cline{2-3}
\rule{0pt}{2.6ex}& Finetuned Diffusion Models & \cite{hu2021lora,gal2022image, ruiz2023dreambooth, trabucco2023effective, trabucco2023effective}  & & & &\\
\hline \hline

\rule{0pt}{2.6ex}\multirow{5}{*}{\textbf{Attack}} & Classifier Based & \cite{hayes2017logan, zhang2024generated, wu2022membership} &  Classifier Based & \cite{hayes2017logan, zhang2024generated}  & Classifier Output Based & \cite{yeom2018privacy, song2021systematic, salem2018ml, song2021systematic, choquette2021label} \\
\cline{2-3} \cline{6-7}
\rule{0pt}{2.6ex}& Distribution Estimation & \cite{hilprecht2019monte,chen2020gan, van2023membership} & & & Likelihood Based & 
\rule{0pt}{2.6ex}\cite{carlini2022membership, zarifzadeh2024low}\\
\cline{2-7}
\rule{0pt}{2.6ex}& Shadow-model & \cite{carlini2022membership} &Distribution Estimation&\cite{hilprecht2019monte,chen2020gan, van2023membership}  & Shadow-model & \cite{shokri2017membership, salem2018ml} \\
\cline{2-3} \cline{6-7}
\rule{0pt}{2.6ex}& Reconstruction & \cite{pang2023black, wu2022membership, chen2020gan} &  & & Gradient Based & \cite{geiping2020inverting, geiping2020inverting, melis2019exploiting}  \\
\cline{2-3}
\rule{0pt}{2.6ex}& Diffusion Process Based & \cite{matsumoto2023membership, Dubinski2023TowardsMR, duan2023diffusion, kong2024an, Hu2023MembershipIO, zhai2024membership} &  &  &  &  \\
\hline \hline

\rule{0pt}{2.6ex}\multirow{5}{*}{\textbf{Mitigation}} & Regularization & \cite{duan2023diffusion, pang2023white, luo2024privacy} & & & Regularization & \cite{wang2020against, kaya2021does, yu2021does, nasr2018machine}
\\
\cline{2-3} \cline{6-7}
\rule{0pt}{2.6ex}& Differential Privacy & \cite{abadi2016deep,torkzadehmahani2019dp, chen2020gs, bie2023private, jiang2022dp, dockhorn2022differentially, ghalebikesabi2023differentially, chen2022dpgen, wang2024dp, cao2021don}     &  Differential Privacy & \cite{dwork2006calibrating, fung2010privacy, zheng2023differentially, harder2021dp}  & Differential Privacy & \cite{abadi2016deep, jayaraman2019evaluating, choquette2021label}
 \\
\cline{2-3} \cline{6-7}
\rule{0pt}{2.6ex}& Knowledge Distillation & \cite{long2021g, papernot2018scalable, fernandez2023privacy} &  & &  Knowledge Distillation & \cite{papernot2018scalable, li2023exploring, tang2022mitigating}  \\
\cline{6-7}
\rule{0pt}{2.6ex}&   &  &  & & Target Sample Exclusion & \cite{jarin2022miashield, conti2022vulnerability}  \\
\hline

\end{tabular}
\end{table*}

\section{Survey of Generative Models} \label{sec:generative models}

In this section, We survey various types of generative models and possible attacks and defense methods thereof. 

\subsection{Models}
We first categorize existing methods of synthetic image generation without attacks or mitigation into consideration.
\paragraph{GANs and VAEs.} GANs~\cite{goodfellow2014generative} introduced a framework where a generator and a discriminator are trained simultaneously in a minimax game, which allows the generator to learn to produce data that closely resembles the real datasets. This framework was further extended by CGAN~\cite{mirza2014conditional} and DCGAN~\cite{radford2015unsupervised}, which added conditional information and architectural improvements, respectively. The introduction of PGGAN~\cite{karras2017progressive} revolutionized the field by enabling progressively growing resolutions, followed by advancements in methods~\cite{zhang2019self, brock2018large, salimans2018improving, karras2020training}, which further improved image fidelity and model control. In parallel, VAEs~\cite{kingma2013auto} use a probabilistic framework to encode data into a latent space and reconstruct it, though they suffer from blurry outputs. Subsequent improvements~\cite{chen2016variational, van2017neural, razavi2019generating} introduced more sophisticated decoding techniques and discrete latent variables to enhance the image quality and make VAEs competitive with GANs in generative tasks. Several works~\cite{genevay2018learning, patrini2020sinkhorn, deja2020end} adopt optimal transport for more effective generator training by incorporating methods such as Wasserstein distance and Sinkhorn divergences to improve the stability and efficiency of the optimization process.

\paragraph{Finetuned GAN/VAE models.} Finetuning techniques for GANs and VAEs have been proposed to improve their adaptability to new tasks and datasets with limited training. FreezeD~\cite{mo2020freeze} fine-tunes GANs by freezing early layers in the discriminator to enhance efficiency and preserve pre-learned features while improving generation quality for new datasets. Similarly, Karras et al.~\cite{karras2020training} introduce adaptive discriminator augmentation, which selectively applies augmentations during GAN training to prevent overfitting in limited data scenarios. These approaches ensure stable and high-quality image generation by leveraging pre-trained models.

\paragraph{Diffusion Models.} Diffusion models have recently gained attention for their ability to generate high-quality images through iterative denoising processes. DDPM~\cite{ho2020denoising} surpasses traditional models such as GANs by reversing a diffusion process to generate photorealistic images. DDIM~\cite{song2020denoising} improved upon DDPM by introducing deterministic sampling to speed up the generation process while maintaining image quality. In parallel, score-based generation methods~\cite{song2019generative, song2020improved, song2020score, dhariwal2021diffusion, batzolis2021conditional} use score matching to estimate the gradient of the data distribution at each noise level. Latent Diffusion~\cite{rombach2022high} further refined diffusion models by operating in a compressed latent space to reduce computational costs and enable high-resolution image generation. These models have been widely applied to various image generation tasks, such as image inpainting~\cite{lugmayr2022repaintinpaintingusingdenoising, xie2023smartbrush, yang2023uni}, editing~\cite{meng2021sdedit, brooks2023instructpix2pix, kawar2023imagic, hertz2022prompt, zhang2023adding, mokady2023null, shi2024dragdiffusion}, and text-guided image generation~\cite{saharia2022photorealistic} due to their ability to handle conditioning. 

\paragraph{Finetuned Diffusion Models.} Following the introduction of LoRA~\cite{hu2021lora}, numerous methods have been developed to enhance diffusion models further. Textual Inversion~\cite{gal2022image} involves finding the most representative text embeddings through inversion techniques to generate high-quality synthetic images. Dreambooth~\cite{ruiz2023dreambooth} addresses the challenge of preventing forgetting by using prior preservation loss during fine-tuning. In addition to these fine-tuning approaches, several works have leveraged Stable Diffusion's pre-trained knowledge for downstream tasks~\cite{trabucco2023effective, azizi2023synthetic, he2022synthetic, shipard2023diversity, bansal2023leaving, zhu2024distribution}. Methods including DA-Fusion~\cite{trabucco2023effective} and Diffusion Inversion~\cite{zhou2023training} further explore synthetic image generation through inversion applied to text embeddings and conditional networks, respectively, to highlight the growing interest in leveraging diffusion models for classification tasks.

\subsection{Attacks}\label{sec:generative-model-attacks}

In recent years, many MIA methods have been proposed for generative models, each tailored to specific models or operational contexts. For instance, Wu et al.~\cite{wu2022membership} is specifically designed for text-to-image diffusion models, while Pang et al.~\cite{pang2023white} is applicable solely in white-box settings. Despite the diversity in these methods, they are underpinned by common conceptual frameworks. Here, we categorize all MIAs into five fundamental types, transcending the specific requirements for model access:

\paragraph{Classifier Based.} This method involves training a binary classifier using labeled data. When the training dataset $D_{aux}^{m}$ is accessible, it is labeled as a member; otherwise, synthetic data $D_{syn}$ is used as member data. Non-member data is represented by auxiliary nonmember data $D_{aux}^{nm}$. Representative methods include LOGAN \cite{hayes2017logan}, \cite{zhang2024generated}, and \cite{wu2022membership} Attack 1. This approach is applicable in white, grey, and black-box scenarios. For instance, a GAN's discriminator can serve as the classifier if the model is accessible \cite{hayes2017logan}.

\paragraph{Distribution Estimation Based.} This method estimates the distribution around the query sample $x_q$. Direct estimation of the generated distribution $p_G$ can be performed using methods such as MC \cite{hilprecht2019monte} and black-box GAN-Leaks \cite{chen2020gan}. 

\paragraph{Reconstruction Based.} This method leverages the generator $g$ to create a reconstruction $x_q'$ of the query sample $x_q$. It assesses membership by comparing the distance or semantic similarity between $x_q$ and $x_q'$. The reconstruction process can be either conditional or unconditional. In the conditional setting, the generator uses inputs such as a text prompt or the class label of $x_q$ to generate $x_q'$ (\emph{e.g.}, \cite{pang2023black} Attacks 1-4, \cite{wu2022membership} Attacks 2-3). Conversely, in the unconditional setting, the generator operates solely on a latent code $z$, with techniques including grey-box GAN-Leaks \cite{chen2020gan} sampling different latent codes to more accurately mimic $x_q$.
    
\paragraph{Shadow-model Based.} LiRA \cite{carlini2022membership}  trains multiple shadow models to estimate likelihood ratios. This method generates models under different hypotheses—specifically, models trained with and without a particular data record—to simulate the target model's behavior on known versus unknown data.

\paragraph{Diffusion Process Based.} MIA methods in this category are specifically designed for the architecture of diffusion models. In white- or grey-box settings, the adversary can access internal components during the diffusion process such as the loss (\cite{matsumoto2023membership, Dubinski2023TowardsMR}, SecMI \cite{duan2023diffusion}, PIA \cite{kong2024an}, \cite{Hu2023MembershipIO} Attack 1), likelihood (\cite{Hu2023MembershipIO} Attack 2, CLiD\cite{zhai2024membership}), or gradients (GSA \cite{pang2023white}) at each time step $t$ to conduct attacks.

However, it is important to note that the performance of an attack—and consequently the evaluated privacy risk—is not directly comparable across different MIA approaches due to varying assumptions about generative models and threat models.

\subsection{Mitigations}\label{sec:generative-model-mitigation}

The mitigation methods of generative models can be categorized into three major approaches: Regularization, DP, and Knowledge Distillation. 

\paragraph{Regularization.} Regularization strategies are designed to reduce overfitting and improve privacy by making models less dependent on specific training data. In generative models, regularization such as Cutout and Random Horizontal Flip have been shown to help. generalize models and minimize the risks of MIAs~\cite{duan2023diffusion, pang2023white}. These augmentations enhance robustness by preventing over-reliance on individual data points. In addition, SMP-LoRA~\cite{luo2024privacy} employs adversarial training to optimize the balance between privacy and utility for safeguarding the model from MIAs. 

\paragraph{DP.} DP remains one of the most effective techniques for ensuring privacy in generative models. DP-SGD~\cite{abadi2016deep} is a core method applied during training to limit the influence of individual training samples by adding noise to gradients. GAN-based models including DP-CGAN~\cite{torkzadehmahani2019dp}, GS-WGAN~\cite{chen2020gs}, and DPGANr~\cite{bie2023private} have integrated DP to ensure the generated data remains private. DP techniques have also extended to VAEs with DP-VAE~\cite{jiang2022dp}. In diffusion models, DPDM~\cite{dockhorn2022differentially} and DP-diffusion ~\cite{ghalebikesabi2023differentially} introduce the training recipes for integrating DP-SGD into unconditional diffusion model training, while DP-LDMs\cite{lyu2023differentially} focus on attention modules to improve the utility-privacy tradeoff in latent diffusion models. In energy-based models, DPGEN\cite{chen2022dpgen} and DP-Promise\cite{wang2024dp} combine DP with energy-based models to generate synthetic data with privacy protection. Furthermore, DP-Sinkhorn~\cite{cao2021don} Sinkhorn divergence to align real and synthetic data for ensuring privacy-preserving synthetic generation. 

\paragraph{Knowledge Distillation.} G-PATE~\cite{long2021g} extends the PATE~\cite{papernot2018scalable} framework to generative models, which use multiple teacher models trained on disjoint subsets of data to guide a student model in producing privacy-preserving synthetic data. This method leverages noise addition during teacher-student training to ensure DP. Privacy distillation~\cite{fernandez2023privacy} introduces a method in which student models are trained on filtered synthetic images produced by a teacher model. This method preserves privacy by leveraging knowledge distillation and excluding high-risk samples during the distillation process. It effectively minimizes the risk of information leakage while maintaining model performance.

\section{Survey of Synthetic Image Release} \label{sec:synthetic images}

Once a generative model is trained on the private dataset, it can produce a synthetic dataset $D_{syn}$ by sampling. 
Releasing this synthetic data, however, still carries inherent privacy risks. In this section, we survey MIA attacks and defenses when $D_{syn}$ is released to adversaries.

\subsection{Models}

Various generative models in the previous section can be used to generate a static dataset to be released in public, which in turn will be used to train a classifier for target variables of interest. 
Furthermore, there are methods of learning synthetic images directly from the training images without learning a generative model first, such as DC (data condensation). DC methods~\cite{wang2018dataset, zhao2020dataset, nguyen2020dataset, zhao2021dataset, nguyen2021dataset, cazenavette2022dataset, zhao2023dataset} aim to synthesize compact and informative datasets that can achieve competitive performance comparable to full real training datasets. Rather than utilizing the entire dataset, these methods condense essential information into smaller synthetic datasets that retain the statistical properties of the original data, thereby reducing training time and computational costs. Zhao et al.~\cite{zhao2020dataset} optimize synthetic datasets by matching the optimization trajectories of models trained on synthetic and real data. Cazenavette et al.~\cite{cazenavette2022dataset} focuses on maintaining the parameter trajectory of model training. Recently, Zhao et al.~\cite{zhao2023dataset} leverages Maximum Mean Discrepancy (MMD) for aligning real and synthetic data in the latent space of randomly initialized ConvNet. These methods offer solutions for efficiently training machine learning models by reducing the dataset size without sacrificing performance.

\subsection{Attacks}\label{sec:image-attacks}
In scenarios where only a fixed size of synthetic data \(D_{syn}\) is released and the generator is inaccessible—even through an API— not all black-box MIAs are viable. Specifically, only \textbf{Classifier Based} and \textbf{Distribution Estimation Based} MIAs can be effectively implemented in this restricted setting:

\paragraph{Classifier Based.}  LOGAN \cite{hayes2017logan} and Zhang et al. \cite{zhang2024generated} both utilize the strategy of training a binary classifier to differentiate between member and non-member data by treating synthetic images as member data. While \cite{zhang2024generated} allows for non-member auxiliary data $D_{aux}^{nm}$ to be sampled from distributions close to the original, LOGAN adopts a stronger attack, assuming that adversaries can directly access non-member data sampled from the original distribution.

\paragraph{Distribution Estimation Based.} Methods such as MC \cite{hilprecht2019monte} and GAN-Leaks \cite{chen2020gan} focus on estimating the generated distribution around the query data $x_q$. Specifically, MC \cite{hilprecht2019monte} counts the proportion of generated points within a predefined neighborhood around $x_q$ to approximate the generated distribution at that point. 
And GAN-Leaks \cite{chen2020gan} utilizes $D_{syn}$ to approximate the probability by measuring the distance between $x_q$ and its nearest neighbor in $D_{syn}$. 
On the other hand, methods such as DOMIAS \cite{van2023membership} estimate both the generated and training distributions.
This approach involves accessing both $D_{syn}$ and $D_{aux}^{nm}$ and employs density estimators to approximate the generator's output distribution as well as the underlying training distribution.



\subsection{Mitigations}\label{sec:image-mitigation}

Publishing synthetic data from models such as GANs or diffusion models may initially seem safe, as the data does not directly represent real individuals. However, synthetic data can still be vulnerable to attacks (\emph{e.g.}, GAN-Leaks and DOMIAS), which can infer membership about the real data used during training. Historically, adding noise directly to data was one of the earliest approaches~\cite{dwork2006calibrating, fung2010privacy} to creating privacy-preserving synthetic data, but it had limited effectiveness. DP-MERF~\cite{harder2021dp} leverage random Fourier features for synthetic image optimization. DPDC~\cite{zheng2023differentially} tackles this by introducing noise during the data condensation process. DPDC implements noise addition during the condensation training phase or directly onto the synthetic datasets after the condensation to ensure privacy preservation while minimizing the risk of privacy attacks.  

\section{Survey of Synthetic Data Trained Classifiers} \label{sec:classifiers}
In this section, we focus on classifier $f$ trained using synthetic data $D_{syn}$ to examine the potential privacy risks when such classifiers are released to adversaries. Although recent studies have explored the utility of synthetic data in replacing or augmenting real datasets for classifier training~\cite{he2022synthetic, azizi2023synthetic, trabucco2023effective, zhou2023training, yuan2023real, askari2023feedback}, the privacy issue has not been addressed carefully. When classifiers trained on synthetic data are shared, they may still reveal sensitive information about the original data distribution. This survey focuses on MIA attacks and defenses when $f$ is released to adversaries.

\subsection{Models}

In this stage, a classifier $f$ with parameter $\theta_f$ is trained with $D_{syn}$ by minimizing the classification loss:
\[
    \min_{\theta_f}\;\; \mathbb{E}_{(\hat{x}, y) \sim {D}_{syn}} \left[ \mathcal{L}(f(\hat{x}; \theta_f), y) \right].   
\]
Any model can be used for the classifier $f$, including large pre-trained models trained from a dataset disjoint from $D_{tr}$ or $D_{syn}$.

\subsection{Attacks}\label{sec:classifier-attacks}

MIAs on classifiers have been extensively studied in both black-box and white-box settings. These attacks typically fall into four categories: Classifier Output Based, Shadow-model Based, Likelihood Based, and Gradient-Based MIAs.

\paragraph{Classifier Output Based.} This approach usually operates under black-box settings where adversaries rely on information such as loss, confidence score, or entropy to infer membership. Yeom et al.~\cite{yeom2018privacy} introduce classifier's loss-based attacks as a key approach, where the loss computed based on the model's output is used to determine whether a data point was part of the training set. Song et al.~\cite{song2021systematic} further improve loss-based attacks by introducing class-wise thresholds. Additionally, Salem et al.~\cite{salem2018ml} exploit the model's prediction vectors, motivated by the observation that overconfident predictions may indicate membership. Song et al.~\cite{song2021systematic} follow a similar idea, as lower entropy in predictions may indicate a higher likelihood of membership. In the more restricted scenario where only predicted labels are accessible to adversaries, Choquette et al.~\cite{choquette2021label} utilize multiple queries to predict membership inference with labels only.

\paragraph{Shadow-model Based.} This method leverages black-box access to mimic the target classifier's behavior by using multiple shadow models. Shokri et al.~\cite{shokri2017membership} propose training shadow models on shadow datasets designed to replicate the target model's training distribution. By observing how the shadow models predict membership, adversaries build a binary classifier that distinguishes between member and non-member data points based on prediction patterns. Salem et al.~\cite{salem2018ml} further simplify this process, requiring less knowledge about the target model's architecture to reduce computational overhead while offering competitive attack efficacy.

\paragraph{Likelihood Based.} This method uses probabilistic approaches to determine membership. LiRA~\cite{carlini2022membership} calculates a likelihood ratio between two hypotheses: one where the data point is part of the training set and another where it is not. This method can be highly effective but is computationally expensive as it requires shadow models for detailed probability estimation across the dataset. RMIA~\cite{zarifzadeh2024low} introduces a more efficient approach that reduces the cost of MIAs by leveraging rejection sampling for likelihood estimation.

\paragraph{Gradient Based.} This approach is highly effective in white-box settings, where adversaries have access to the internal parameters of the target model. Nasr et al.~\cite{nasr2019comprehensive} pioneered this approach by demonstrating that model gradients can reveal sensitive information about the training data. Geiping et al.~\cite{geiping2020inverting} applied gradient inversion methods to recreate entire images from model updates in federated learning settings. Similarly, Melis et al.\cite{melis2019exploiting} highlight that gradient updates in collaborative learning scenarios can leak unintended features. These works emphasize the risks posed by gradient sharing and underscore the importance of developing more robust defenses to mitigate privacy attacks.    

\subsection{Mitigations}\label{sec:classifier-mitigation}
Releasing classifiers trained on synthetic data can still pose privacy risks, especially when generators are overfitted to specific real data points. We can categorize mitigation methods for classifiers into four types: 

\paragraph{Regularization.} Several studies have shown that regularization such as pruning~\cite{wang2020against} and data augmentation~\cite{kaya2021does, yu2021does} reduce overfitting and mitigate MIAs by making models less confident in predictions. These techniques improve the generalization of models by preventing them from becoming tailored to the training data, thus lowering privacy risks. In addition, Adversarial regularization~\cite{nasr2018machine} introduces min-max optimization that directly strengthens models against MIAs by balancing classification accuracy and privacy.

\paragraph{DP.} DP-SGD~\cite{abadi2016deep} ensures that the gradients updated during training do not reveal sensitive information about the underlying training data. Further works~\cite{jayaraman2019evaluating, choquette2021label} have demonstrated how DP-SGD provides privacy guarantees in various scenarios. However, a common tradeoff with DP is the reduction in utility, since models trained with DP often experience a drop in accuracy due to noisy gradients.

\paragraph{Knowledge Distillation.} PATE~\cite{papernot2018scalable} ensures privacy by training multiple teacher models on disjoint subsets of data and aggregating their noisy outputs to train a student model. This approach provides strong DP guarantees by adding noise during the aggregation process. PrompPATE~\cite{li2023exploring} explores the efficiency of visual prompting in a similar framework by adopting a training approach in combination with visual prompts to further enhance privacy efficiency in visual tasks. SELENA~\cite{tang2022mitigating} adopts self-distillation and ensemble learning approaches. They train multiple sub-models on overlapping subsets of data, followed by self-distillation where the outputs of the sub-models are used to train a final model and reduce the risk of MIAs while maintaining classification performance without adding noise to gradients. 

\paragraph{Target Sample Exclusion.} Jarin et al.~\cite{jarin2022miashield} identifies high-risk predictions that are more likely to reveal sensitive data and excludes them from the output before ensembling. This method essentially filters out data points that could be vulnerable to MIAs by providing an additional layer of protection. Conti et al.~\cite{conti2022vulnerability} introduces metrics to assess the vulnerability of data points under multiple MIAs and target models. They show that vulnerability is influenced by the specific attack and the target model, rather than being an inherent characteristic of the data point itself.

\section{Evaluation} \label{sec:benchmark}
The benchmark provides a systematic way to assess privacy risks and utility tradeoffs across generative models and sharing strategies. Our benchmark focuses on the privacy risks associated with sharing synthetic images and classifiers trained on synthetic data. This choice aligns with our aim to use model-agnostic MIAs, which allow for empirical measurement of privacy leakage without the constraints of model-specific assumptions.

We prioritize the utility-privacy tradeoff in a real-world data-sharing context by excluding considerations of specific DP budgets or privacy guarantees. Instead, we use simple and model-agnostic MIA methods as empirical privacy measures instead of SOTA techniques. This choice is based on three reasons. First, simplicity ensures the method is widely applicable and easy to interpret across different models and settings. Second, it highlights fundamental privacy risks without relying on specific attack strategies or model architectures. Finally, using MIAs allows us to directly compare the privacy risks of DP and non-DP approaches. 

We use classifier performance (\emph{e.g.}, accuracy or F1 score) as the utility measure because it provides a direct comparison between the two sharing strategies: (1) sharing synthetic images and (2) sharing classifiers trained on synthetic data. In addition, we exclude FID for synthetic image quality measurements due to the smaller size of our datasets. FID requires large sample sizes to produce reliable estimates and be unstable or unrepresentative on smaller datasets~\cite{heusel2017gans}. Instead, classification accuracy serves as a robust metric for evaluating the utility of synthetic data.

\subsection{Dataset}
To evaluate the utility-privacy tradeoffs, we conduct experiments on three datasets: CelebA~\cite{liu2015faceattributes},  Fitzpatrick 17k~\cite{groh2021evaluating}, and Chexpert~\cite{irvin2019chexpert}. 
A dataset is widely used to test generative models’ performance. For this study, we constructed a subset of CelebA, which consists of 2000 images from 250 randomly selected identities with 8 images per identity. This subset allows us to observe the impact of repeated identity exposure in model training and assess MIA resilience. 
The subset utilizes three key attributes (Male, Smiling, Young), which enables attribute-based evaluation of privacy and utility outcomes. 

For medical dataset evaluation, we use a subset of Fitzpatrick17k dataset, which contains skin disease images. This subset includes 375 images of basal cell carcinoma, 192 of melanoma, and 357 of squamous cell carcinoma skin lesions. The Fitzpatrick17k dataset allows us to examine privacy and utility tradeoffs in privacy-sensitive healthcare domains, where data scarcity can exacerbate privacy risks. 
To further assess generalizability across different medical imaging domains, we include CheXpert, a large publicly available dataset of chest X-ray images. For our experiments, we curated a balanced subset of 2000 images in total, distributed across four key classes: ``No Finding'', ``Cardiomegaly'', ``Pleural Effusion'', and ``Pneumonia'', with 500 images per class. The inclusion of CheXpert allows us to evaluate the utility-privacy tradeoffs in another sensitive healthcare context with distinct visual characteristics.
Across datasets, we further divide the data to ensure consistency in evaluating generative models and membership inference attacks. We adopt the subset auxiliary assumption to assess privacy risks in challenging conditions for evaluating the robustness of synthetic data when an adversary has partial access to the training data. The training set for generative models ($D_{tr}$) combines member auxiliary and member test data ($D_{tr} = D^{m}_{aux} \cup D^{m}_{te}$), serving as the source data for model training. We define an auxiliary dataset ($D_{aux} = D^{m}_{aux} \cup D^{nm}_{aux}$) to simulate potential information an adversary could access, which may include a mix of members and non-members. Finally, the test set ($D_{te} = D^{m}_{te} \cup D^{nm}_{te}$) is used solely for evaluating the success of MIAs.

\subsection{Representative methods}

To evaluate utility-privacy tradeoffs, we benchmark a selection of methods that represent a diverse range of current approaches. Our choices aim to cover various key dimensions in generative modeling and privacy preservation:
\begin{itemize}
\item \textbf{Core Generative Architecutres}: We include prominent model families such as GAN (FreezeD)~\cite{mo2020freeze}, VAE~\cite{kingma2013auto}, and DDPM~\cite{ho2020denoising}. This allows us to assess how different foundational architectures perform.
\item \textbf{Training and Adaptation Strategies}: We examine both models trained from scratch (\emph{e.g.}, VAE, DDPM) and those utilizing fine-tuning methods on pre-trained models (\emph{e.g.}, LoRA~\cite{hu2021lora}, Textual Inversion~\cite{gal2022image}). This distinction is crucial as fine-tuning is a common paradigm for adapting large foundation models.
\item \textbf{Optimization Focus}: The selected methods encompass different optimization goals, including direct image optimization (\emph{e.g.}, DCDM~\cite{zhao2023dataset} and generative model optimization (\emph{e.g.}, GAN, DDPM).
\item \textbf{Privacy Mitigation Approaches}: We compare non-mitigated methods with various privacy-enhancing techniques. These include methods applying DP at different stages or with different mechanisms (\emph{e.g.}, DP-Sinkhorn~\cite{cao2021don} for generators, DP-LDMs~\cite{lyu2023differentially} for attention modules) and adversarial training aimed at mitigating privacy risks (\emph{e.g.}, SMP-LoRA~\cite{luo2024privacy}).
\end{itemize}    
Detailed descriptions of each method are provided in the Appendix~\ref{sec:appendix-methods-details}.

In addition to the above methods, we report the baseline result of \textbf{a classifier trained on real data} with SGD or DP-SGD~\cite{abadi2016deep} with various DP budgets. These classifiers allow us to assess the utility-privacy tradeoffs in comparison to classifiers trained on synthetic data.

\paragraph{Implementation Notes.}
We adopt model-specific training strategies tailored to various generative models. We employ Stable Diffusion v1.5 as the base model for latent diffusion models such as LoRA, DP-LDMs, SMP-LoRA, and TI. BLIP2-generated captions~\cite{li2023blip} are used for CelebA, while disease names in text serve as conditioning for Fitzpatrick17k. We conduct training from scratch for DDPM and VAE models, whereas we utilize a pre-trained StyleGAN2~\cite{karras2020analyzing} model on the FFHQ dataset as the baseline for GAN. For DP-Sinkhorn, we pre-train a VAE generator on a disjoint portion of the CelebA and Fitzpatrick17k datasets to prevent overlap with the training and evaluation data. Except for latent diffusion-based models, conditional training utilizes three attributes (Male, Smiling, and Young) for CelebA and three diseases (basal cell carcinoma, melanoma, and squamous cell carcinoma) for Fitzpatrick17k. 

For each generative method, we identify five key hyperparameters (\emph{e.g.}, DP budget, LoRA rank, or adversarial weight) and systematically explore them to ensure a fair comparison. We use two-fold cross-validation for both CelebA and Fitzpatrick17k, averaging the results of these two folds. Our goal is to find the parameter settings that yield the best tradeoff between utility (\emph{e.g.}, classification accuracy) and privacy (\emph{e.g.}, MIA performance). Further details about these hyperparameters are provided in \emph{Appendix~\ref{sec:appendix-hparams}}.

For each method, we generate three synthetic images per real image. Latent diffusion approaches (LoRA, TI, DP-LDMs, SMP-LoRA) use captions or disease names as conditioning; other methods rely on attributes or disease labels. For DCDM, we experiment with multiple ratios (50\%\,--\,90\%) of the original training size, since its objective is dataset compression rather than conventional generative modeling.

We train a ResNet-18 (pre-trained on ImageNet) on either real or synthetic images. For CelebA, we classify three binary attributes and for Fitzpatrick17k, we classify among three disease categories. To benchmark privacy against real-data baselines, we also compare with classifiers trained using DP-SGD~\cite{abadi2016deep} on the real dataset at varying $(\epsilon, \delta)$ budgets. In particular, we set $\epsilon=\{10,50,100\}$ for DP-SGD as a baseline, and $\epsilon=\{10,50,100,250,500\}$ for DP-LDMs and DP-Sinkhorn to examine a wider utility-privacy spectrum. We fix $\delta$ according to each dataset’s size.

\subsection{Evaluation Metrics}

\paragraph{Utility: Classification Performance.}

We evaluate the utility of synthetic images by replacing them in classifier training. Synthetic images are generated using conditions from the member datasets. A classifier is trained on these synthetic images for CelebA attribute classification (\emph{e.g.}, male, smiling, young) and Fitzpatrick17k skin disease classification. The performance of this classifier is then compared to one trained on the real dataset to assess the utility of the synthetic data.

\paragraph{Privacy: Attack Success Rate.}

We evaluate privacy risks by measuring the TRR at low FPR and Area under the curve (AUC) in two types of MIAs. \textbf{Attack 1} assesses privacy when synthetic images are made public by comparing similarities between real and synthetic images in the latent space. These attacks help determine the level of privacy risk associated with either classifiers or synthetic data. \textbf{Attack 2} evaluates privacy using a loss-based approach to infer membership when the classifier is made public.

We choose these two attacks to provide a universal privacy assessment of diverse methods. Attack 1 enables model-agnostic analysis regardless of how the images are generated. Attack 2 also enables classifier-agnostic analysis for any classifier with a loss function. We do not directly attack generative models because most existing attacks on generative models make specific assumptions about the model architecture, training processes, or require access to internal components, which makes it difficult to compare different methods fairly.

\paragraph{Attack 1: Synthetic Image-based Attack.}

We assume the adversary is provided with synthetic images from $G$ and auxiliary datasets of members and non-members $D_{aux} = D^m_{aux}\cup D^{nm}_{aux}$. Inspired by Pang et al.'s approach~\cite{pang2023black}, we calculate the similarity between real query images and synthetic data in the latent space. However, our attack differs in that we compare each query image against the entire synthetic dataset rather than only the corresponding synthetic images as in Pang et al.'s approach. This adjustment makes the attack generator agnostic, which allows it to work with methods such as DC that cannot perform conditional generation.
We use a pre-trained encoder to extract feature embeddings for each query image $x$ and synthetic dataset $D_{syn}$. For each query image, we compute its similarity against all images in the synthetic dataset and select the top $k$ similarity scores to obtain a robust similarity measure. The averaged similarity is then fed into a classifier to predict membership. This attack is based on the assumption that generators tend to produce synthetic images that closely resemble the training set. By leveraging this characteristic, we can infer membership by identifying synthetic images that are highly similar to specific query images.

\paragraph{Attack 2: Loss-based Classifier Attack.}
We assume the adversary has black-box access to classifiers $f$ and examples of members and non-members auxiliary datasets $D_{aux} = D^m_{aux}\cup D^{nm}_{aux}$. The adversary targets real data membership inference against classifiers trained on synthetic data, following the loss-based method~\cite{yeom2018privacy}. The adversary computes loss values $\ell(x, y)$ for samples in the auxiliary set $D_{aux}$ using the trained classifier $f$. An attack classifier $A$ is used on these losses to predict membership labels. The attack is based on the assumption that a classifier's loss will be lower for training samples or synthetic data that reflects sensitive information from the original set. We can infer membership by identifying samples that yield lower losses.

\paragraph{Attack 3: Individual Risk via LiRA with Mislabeling.}
To assess privacy risks at an individual level, moving beyond population-level metrics, and inspired by recent critiques of standard MIA evaluations~\cite{aerni2024evaluations}, we implement a third attack based on the Likelihood Ratio Attack (LiRA) framework~\cite{carlini2022membership}. This attack targets \textbf{individual-level risk}. For this, we train a set of shadow models. During the training of these shadow models, a specific subset of training samples (\emph{e.g.}, 100 samples serving as canaries for evaluation) is \textbf{intentionally mislabeled} (i.e., their labels are flipped) for some shadow models, while their original, correct labels are used for training other shadow models. By comparing the shadow model outputs (\emph{e.g.}, confidence scores) on these specific canary samples under both the correct and intentionally flipped labeling hypotheses, we compute likelihood ratios to infer their membership. This allows for deriving individual membership scores and evaluating metrics such as TRP at low FPR using sample-specific thresholds to provide a more robust measure of individual privacy leakage.

\subsection{Results}

\begin{figure*}[ht]
    \centering
    \includegraphics[width=0.84\textwidth]{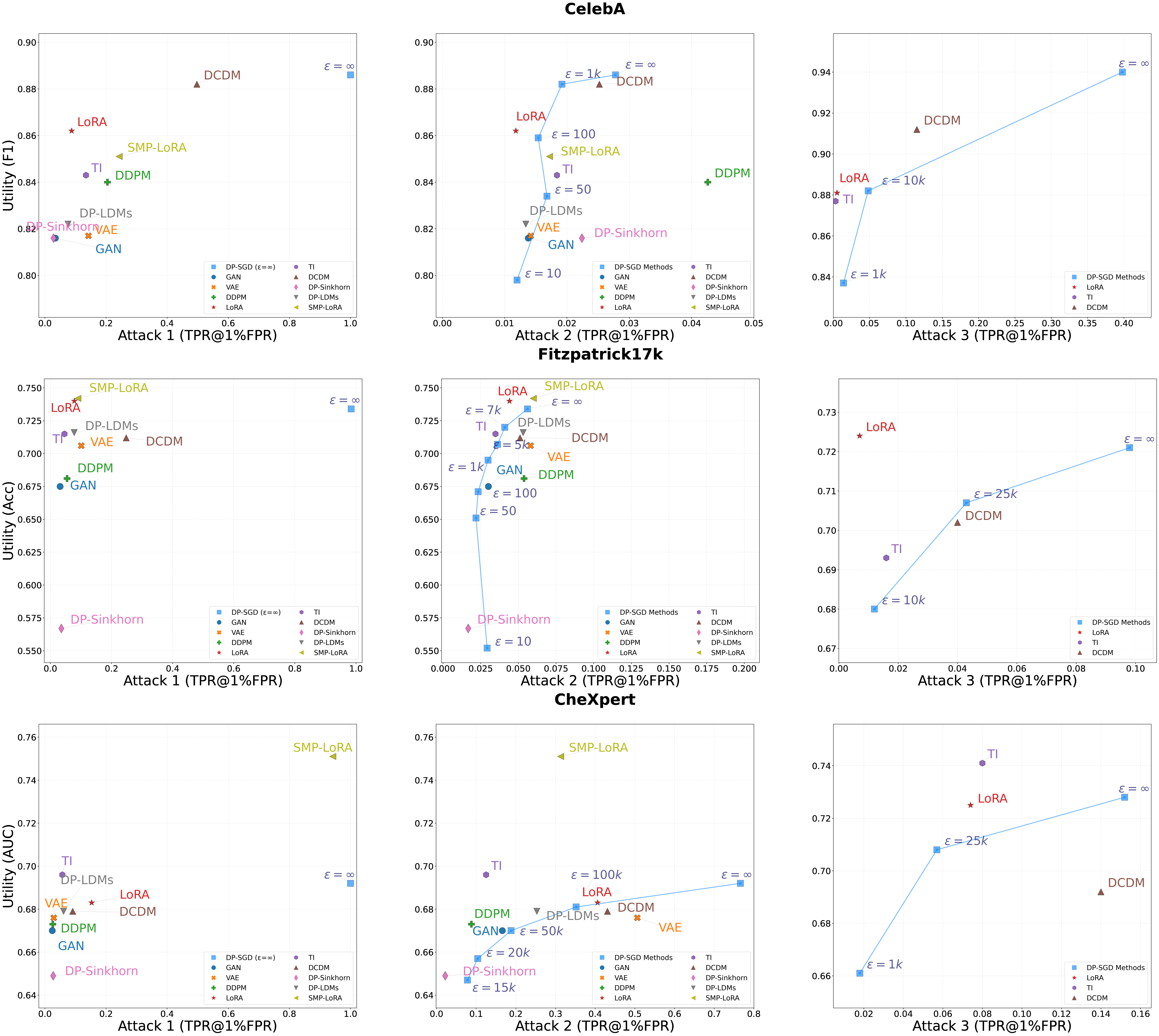}
    \caption{Privacy-Utility Tradeoff on CelebA (Top), Fitzpatrick17k (Middle), and CheXpert (Bottom). Each row displays results for Attack 1 (left, synthetic data release) and Attack 2 (middle, classifier release), and Attack 3 (right, classifier release focusing on individual-level risk). The x-axis represents privacy risk (TPR@1\%FPR, lower is better), while the y-axis indicates utility (dataset-specific: F1-Score for CelebA, Accuracy for Fitzpatrick17k, AUC for CheXpert; higher is better). Optimal methods are located in the top-left of each subplot. Note: X-axis ranges vary across attacks for improved visibility of method distribution.}
    \label{fig:results}
\end{figure*}

We answer the following research questions with the proposed benchmarking results.

\paragraph{RQ1. Utility-privacy tradeoff of sharing synthetic images vs. sharing classifiers trained on synthetic data.}

Figure \ref{fig:results} demonstrates that the optimal data-sharing strategy—releasing synthetic images directly (Attack 1) versus releasing classifiers trained on them (Attack 2) is dataset-dependent. For CelebA, where its visually unique images, sharing classifiers consistently yields superior privacy (lower Attack 1 TPR vs. Attack 2 TPR, \emph{e.g.}, ~0.087 vs. ~0.012). This trend also holds for Fitzpatrick17k (\emph{e.g.}, LoRA: Attack 1 TPR ~0.079 vs. Attack 2 TPR ~0.044). In contrast, for CheXpert, where images share more common features, direct synthetic image sharing (Attack 1) can be less risky for the same methods (\emph{e.g.}, LoRA: Attack 1 TPR ~0.153 vs. Attack 2 TPR ~0.406). This highlights how dataset characteristics influence attack efficacy, a point further explored in Section \ref{sec:dataset_specific_observations}.

\textbf{Takeaway:} The safer strategy for releasing synthetic outputs (images vs. classifiers) varies with dataset characteristics. Classifiers tend to be a safer release mechanism for visually diverse datasets such as CelebA, while direct image sharing may be preferable for certain methods on more homogeneous datasets such as CheXpert. \textbf{Nonetheless, any synthetic data approach offers a fundamental privacy improvement over sharing real images directly.}

\begin{figure*}[ht]
    \centering
    \includegraphics[width=.59\textwidth]{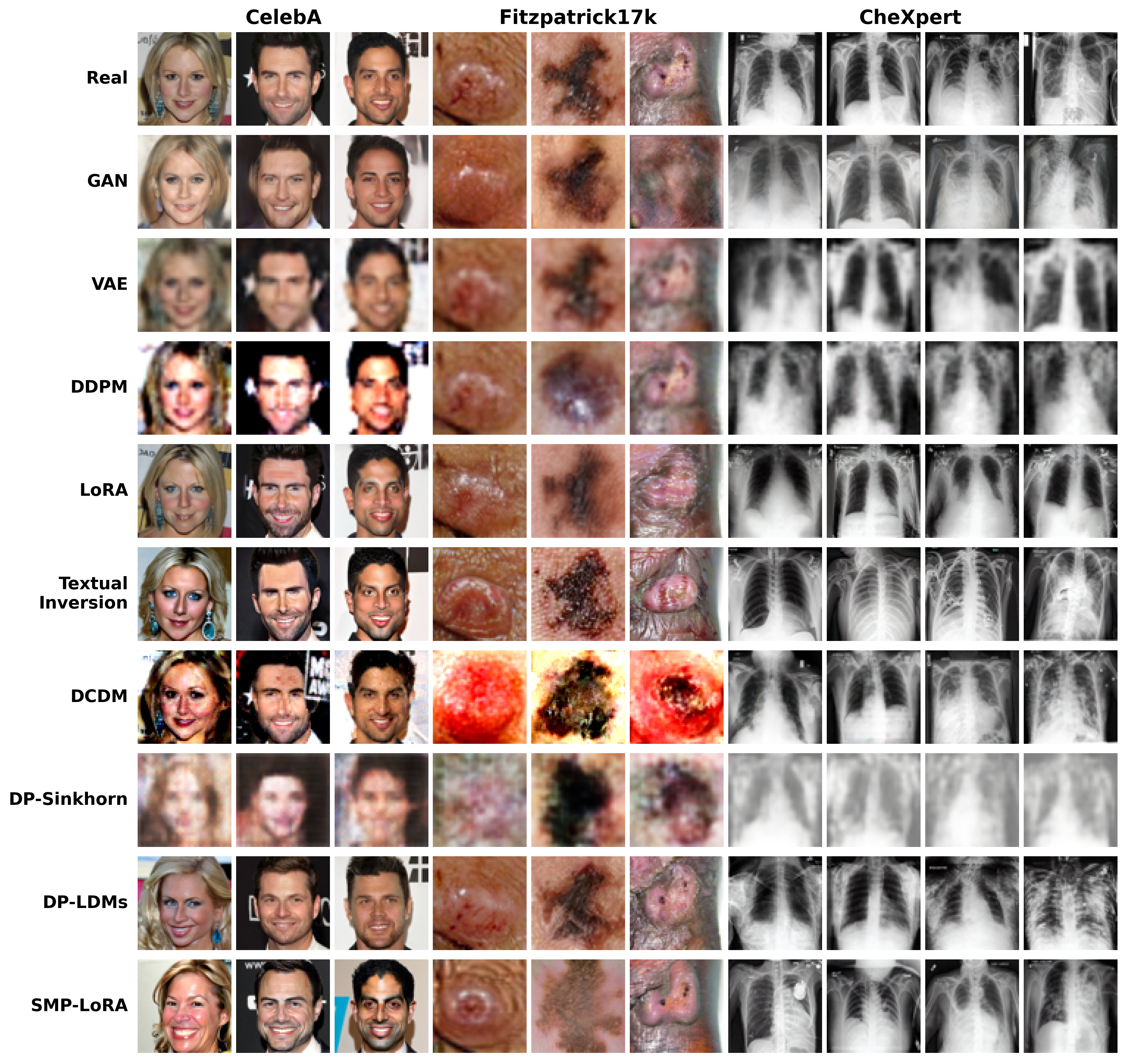}
    \caption{Examples of synthetic images generated by different methods for CelebA (Left), Fitpatrick17k (Middle), and Chexpert (Right). Each row corresponds to a different method, and each column represents specific real images with corresponding synthetic variants.}
    \label{fig:celeba}
\end{figure*}

\paragraph{Dataset-Specific Observations and Impact of Visual Characteristics.}\label{sec:dataset_specific_observations}
The varied efficacy of Attack 1 (on images) versus Attack 2 (on classifiers) across datasets strongly indicates that intrinsic dataset characteristics are critical. CelebA images, with their high inter-subject visual diversity, present many distinct features as shown in Figure \ref{fig:celeba}. If synthetic images replicate these unique training instance details too closely, image-based MIAs (Attack 1) can more readily exploit these similarities. \\
In contrast, CheXpert X-rays often exhibit considerable visual similarity, both within the same class and subtly between different pathological conditions. This general visual homogeneity can diminish the effectiveness of Attack 1 by making it challenging for an attacker to pinpoint unique features of training instances or their synthetic counterparts. Consequently, for some generative methods on CheXpert, Attack 1 may be less potent than classifier-based attacks (Attack 2), which leverage the model's learned decision boundaries. \\
Fitzpatrick17k skin lesion images offer a different challenge. Its defined disease classes serve as targets for classification, but the visual distinctiveness of individual instances is generally less than CelebA's unique faces. Moreover, significant visual feature overlap between these classes (\emph{e.g.}, similar lesion types or color patterns across conditions) is common.
These observations establish that dataset-inherent visual properties—including inter-subject diversity, intra-class variance, and the degree of inter-class feature overlap—are crucial determinants of MIA vulnerability landscapes and significantly influence the relative risk associated with different data sharing strategies and attack vectors.

\paragraph{RQ2. Utility-privacy tradeoff with and without mitigation.}
Mitigation techniques aim to enhance the privacy of generative models (Stage A protection) and subsequent artifacts. Our benchmark (Figure \ref{fig:results}, Tables \ref{tab:celeba}, \ref{tab:fitzpatrick}, \ref{tab:chexpert}) shows that explicit mitigations such as adversarial training (SMP-LoRA) often achieve strong privacy while maintaining high utility, sometimes surpassing the non-private baseline (DP-SGD ($\epsilon=\infty$)) utility on Fitzpatrick17k and CheXpert. DP generative methods such as DP-LDMs and DP-Sinkhorn also enhance privacy. DP-Sinkhorn consistently provide the lowest attack success rates (\emph{e.g.}, Attack 1 TPR for Fitz: 0.036) but often at a significant utility cost (Accuracy for Fitz: 0.567). However, explicit mitigation is not the only path to privacy. Non-mitigated models (\emph{e.g.}, LoRA or TI) through careful hyperparameter tuning (Appendix \ref{sec:appendix-hparams}), such as adjusting LoRA rank, also achieve considerable privacy gains, acting as a form of implicit regularization. This highlights that the choice between explicit and implicit mitigation hinges on the specific utility-privacy goals and acceptable overhead.

\textbf{Takeaway:} Adversarial training (SMP-LoRA) shows notable efficacy. While DP for generative models improves privacy, it can reduce utility more than well-tuned non-mitigated or adversarially trained models, which offers alternative paths to favorable utility-privacy tradeoffs.


\paragraph{RQ3. DP-classifier vs. classifier trained on synthetic images.}
To compare classifiers trained on real data with DP-SGD against classifiers trained on synthetic data (from various generative models), we primarily focus on Attack 2 results (classifier release scenario) in Figure \ref{fig:results}. Across CelebA, Fitzpatrick17k, and CheXpert, several classifiers trained on synthetic data from methods such as SMP-LoRA, LoRA, and TI demonstrate superior utility-privacy trade-offs compared to DP-SGD classifiers trained on real data. For instance, on Fitzpatrick17k, LoRA (Utility: 0.740, Attack 2 TPR: 0.044) and SMP-LoRA (Utility: 0.742, Attack 2 TPR: 0.059) achieve higher utility than the baseline "DP-SGD ($\epsilon=\infty$)" (Utility: 0.734, Attack 2 TPR: 0.056) while offering significantly better privacy. Even compared to weaker DP-SGD settings in real data (\emph{e.g.}, $\epsilon=7e3$), these top synthetic data-trained classifiers often provide a better balance. This underscores the potential of using synthetic data as an intermediary for training downstream models, which can sometimes offer better privacy for a given utility than directly applying DP to the real-data classifier training.

\textbf{Takeaway:} Classifiers trained in high-quality synthetic data (especially from advanced diffusion-based methods) can achieve better utility-privacy trade-offs than classifiers trained directly on real data with DP-SGD.

\paragraph{RQ4. Utility-Privacy Tradeoff of GAN vs DM vs others.}
Examining the performance of different generative model families (Figure \ref{fig:results}), diffusion-based approaches (LoRA, TI, DDPM, DP-LDMs, SMP-LoRA) generally deliver superior utility-privacy trade-offs across both Attack 1 and Attack 2 scenarios compared to GAN and VAE. This suggests diffusion models' capacity for high-fidelity synthesis can translate into useful data that, depending on the specific attack and dataset, may present a better balance.
In addition, the expected trend of "higher utility implies lower privacy" is not always monotonic and reveals "mutant cases" depending on the specific attack vector and dataset. For instance, on CelebA against Attack 2, LoRA (Utility: 0.862, A2 TPR: 0.012) shows higher utility than GAN (Utility: 0.816, A2 TPR: 0.014), yet LoRA also presents better privacy. A similar inversion is seen on Fitzpatrick17k against Attack 2, where LoRA (Utility Acc: 0.740, A2 TPR: 0.044) demonstrates higher utility than VAE (Utility Acc: 0.706, A2 TPR: 0.058) while simultaneously achieving better privacy. These instances indicate that some advanced generative architectures might be inherently more robust against certain types of classifier-based MIAs (Attack 2) even at high utility, or that their failure modes under attack are different.

\textbf{Takeaway:} Diffusion-based methods generally lead to more favorable utility-privacy trade-offs for synthetic data generation compared to the tested GAN and VAE models. In addition, the relationship is not always linear. Some high-utility diffusion models (\emph{e.g.}, LoRA) can exhibit surprisingly strong privacy against specific attacks (\emph{e.g.}, Attack 2) compared to other models with lower utility.


\paragraph{RQ5. Synthetic Image Quality vs. Utility.}
Analyzing the synthetic images (Figures \ref{fig:celeba}) reveals substantial variation. Non-mitigated GAN, VAE, and diffusion-based models (\emph{e.g.}, DDPM, LoRA, and TI) produce realistic images while preserving key features. Among the mitigated methods, DP-LDMs and SMP-LoRA maintain visual realism and essential features. These methods demonstrate that mitigation techniques can retain important class-specific details while offering privacy protection. In contrast, DP-sinkhorn struggles to generate visual coherence, impacting its utility, while DCDM produces some images that appear as blends of multiple samples.

\textbf{Takeaway:} Higher visual quality and feature representation in synthetic images usually correlate with better utility. Diffusion models and some mitigated methods (DP-LDMs, SMP-LoRA) balance fidelity and privacy well, unlike DP-Sinkhorn which sacrifices quality for privacy.

\paragraph{RQ6. Disparate Impact of DP-SGD and Synthetic Data Methods.}
Motivated by findings that DP can reduce accuracy for underrepresented subgroups~\cite{bagdasaryan2019differential}, we analyzed its impact on Fitzpatrick17k by comparing performance on lighter (Fitzpatrick Scale 2, majority) versus darker (Scale 6, minority) skin types (Table \ref{tab:fitz_disparate_impact_analysis}). The non-private baseline (DP-SGD ($\epsilon=\infty$)) established reference accuracies of 0.712 for lighter skin and 0.757 for darker skin. On average, DP-SGD methods with finite $\epsilon$ reduced lighter skin accuracy by 0.069 and darker skin accuracy by 0.090 compared to the baseline, indicating a larger negative impact on the minority skin type. In contrast, synthetic data methods showed a smaller reduction in accuracy from baseline for both groups: -0.049 for lighter skin and -0.021 for darker skin on average. Notably, TI and SMP-LoRA achieved higher utility on minority group.

\textbf{Takeaway:} DP-SGD methods led to a greater reduction in accuracy for both skin types compared to the average of synthetic methods with a greater average drop for the minority skin group. In contrast, synthetic approaches demonstrate a better capacity to maintain or improve performance on specific darker skin types relative to the non-private baseline.

\paragraph{RQ7. Assessing Individual-Level Privacy Risks of Generative Models via LiRA.}
To complement population-level MIAs (Attacks 1 \& 2), we employed LiRA with an individual threshold (Attack 3) to estimate individual-level privacy risks by assessing the memorization of specific training samples. For this analysis, our focus was on three representative non-mitigated generative methods: TI, LoRA, and DCDM, due to the computational cost of LiRA. Our findings from this LiRA assessment revealed a notable trend: both TI and LoRA consistently exhibited lower individual privacy risk scores, suggesting less direct memorization of specific training examples compared against the DP-SGD baseline. We hypothesize that this result for TI and LoRA is significantly influenced by our synthetic data generation strategy. Specifically, by producing three distinct synthetic images for each single real training image when using TI and LoRA, these methods may effectively average out or dilute the model's tendency to memorize highly unique or particularly vulnerable training instances. Therefore, generating multiple synthetic images for each real one might make it harder to detect if specific training data was memorized, even when these TI and LoRA setups didn't use formal mitigation strategies. This capability to generate multiple, varied synthetic instances from a single real example is indeed a general strength of generative models, offering the flexibility to produce larger datasets as needed. However, the precise impact of further increasing the number of synthetic images generated per real sample (e.g., beyond our three-to-one ratio) on these individual-level LiRA risk scores remains an open question and a promising direction for future work.

\textbf{Takeaway:} Non-mitigated fine-tuning methods like TI and LoRA, when paired with a multi-instance synthetic generation strategy (e.g., producing three synthetic images per real image), can surprisingly result in lower individual training sample memorization, as measured by LiRA, compared to a DP-SGD baseline. This suggests that certain generative process designs may inherently offer some mitigation against individual-level privacy risks.

\subsection{Discussions}
As shown in Table~\ref{tab:celeba}, Table~\ref{tab:fitzpatrick}, and Table~\ref{tab:chexpert} (Appendix), hyperparameters play a crucial role in balancing the utility-privacy tradeoff in generative models with settings such as DP budget, LoRA rank, and adversarial loss strength directly influencing the performance. Careful selection of these parameters can help identify Pareto-optimal points where neither privacy nor utility can be improved without compromising the other. 
For instance, adjusting the DP budget in DP-SGD impacts the privacy protection level but can also degrade utility if set too aggressively. Similarly, the LoRA rank affects model complexity and fine-tuning efficiency at the cost of privacy. Likewise, the adversarial loss strength in SMP-LoRA determines the model's tradeoff point. 

Our study opens up several avenues for future research.
First, a critical next step is the systematic optimization of hyperparameters for generative models to delineate the Pareto optimal frontier for utility-privacy tradeoffs. Achieving this, potentially using advanced techniques like Bayesian Optimization \cite{avent2019automatic}, would provide practitioners with a clear set of best-achievable balances between data utility and privacy.
Second, moving beyond data replacement, future work should investigate the use of synthetic data for augmentation purposes within privacy-preserving frameworks. Exploring optimal methods for blending real and synthetic data could significantly improve utility-privacy tradeoffs, especially in data-scarce or highly sensitive domains, thereby expanding our understanding and application of PPDS.

\section{Conclusion} \label{sec:conclusion}
This paper presented a SoK on utility-privacy tradeoffs in synthetic image generation, structured by our generation-classification pipeline. This framework enabled a systematic categorization of methods, attacks, and mitigations, complemented by a comprehensive empirical benchmark using MIAs.
Our SoK reveals that optimal strategies for privacy-preserving data sharing are highly context-dependent, varying significantly with dataset characteristics and model choices. We demonstrated that synthetic data, especially from advanced architectures like diffusion models or when used to train downstream classifiers, can achieve superior utility-privacy balances and potentially fairer outcomes compared to directly applying Differential Privacy to real data. Furthermore, our findings underscore that beyond explicit mitigation techniques, model architecture, specific generation processes (\emph{e.g.}, multi-instance synthesis), and careful hyperparameter tuning critically influence privacy.
Ultimately, this work provides foundational empirical insights and actionable guidance for navigating the varied landscape of privacy-preserving synthetic image generation to steer practitioners towards informed decisions and future research towards promising avenues.

\section{Ethics Considerations}
This research complies with ethical guidelines and ensures privacy by using publicly available, anonymized datasets (CelebA, Fitzpatrick17k). Potential risks of misuse of synthetic data were mitigated by restricting shared artifacts to ethical use under appropriate licensing. The study follows the Menlo Report principles, prioritizing user privacy and safety while balancing utility and privacy in the presented methods.

\section{Open Science}
We adhere to the USENIX open science policy by sharing our source code, processed datasets, and evaluation protocols in a public repository post-review. The source code provided enables the training of all models and the full replication of our experimental results. The raw datasets (\emph{e.g.}, CelebA, Fitzpatrick17k, and CheXpert) are publicly available from their original sources, and our provided documentation includes detailed instructions for accessing them. Our processed data files used for training and evaluation are also included to ensure reproducibility. The complete artifact package is available on Zenodo at the following DOI: \url{https://doi.org/10.5281/zenodo.15605704}.

\section*{Acknowledgments}
This work was supported in part by NSF NAIRR 240230.
\bibliographystyle{plain}
\bibliography{aaai25}

\appendix
\clearpage

\begin{table*}[htbp]
    \centering
    \caption{Hyperparameters and Values Used for Each Method}
    \begin{tabular}{l|l|l}
        \hline
        \textbf{Method} & \textbf{Hyperparameter} & \textbf{Values Used for Tuning} \\ \hline
        VAE & The weight of the KL divergence term & [0.001, 0.0008, 0.0006, 0.0004, 0.0002] \\ \hline
        GAN & Layers to freeze during training & [Last-1, Last-2, Last-3, Last-4, Last-5] \\ \hline
        DDPM & Strength of the SDEdit operation & [0.5, 0.6, 0.7, 0.8, 0.8] \\ \hline
        LoRA & The rank of the LoRA decomposition & [16, 32, 64, 128, 256] \\ \hline
        TI & Classifier-free guidance scale & [1.0, 2.5, 4.0, 5.5, 7.0] \\ \hline
        DCDM & The ratio between real and synthetic images & [50\%, 60\%, 70\%, 80\%, 90\%] \\ \hline
        DP-Sinkhorn & Epsilon values for privacy guarantees & [10, 50, 100, 250, 500] \\ \hline
        DP-LDMs & Epsilon values for privacy guarantees & [10, 50, 100, 250, 500] \\ \hline
        SMP-LoRA & The weight for the adversarial term & [1.0,  0.5, 0.1, 0.05, 0.01] \\ \hline
    \end{tabular}
    \label{tab:hyperparameters_tuning}
\end{table*}

\section{Supplementary Details and Full Results}
\label{sec:appendix-main}

\subsection{Details of Benchmarked Generative Methods}
\label{sec:appendix-methods-details}
In this section, we provide further details on the representative generative methods benchmarked in our study.
\begin{description}
    \item[GAN (FreezeD)~\cite{mo2020freeze}] FreezeD is a GAN-specific fine-tuning method that freezes the layers of the discriminator during training. This prevents overfitting and improves the generalization performance of GAN models, especially when training with limited data.
    \item[VAE~\cite{kingma2013auto}] VAE is a probabilistic generative model that learns to represent data in the latent space. It generates images by sampling from this learned latent distribution and decoding them back to the original image.
    \item[DDPM~\cite{ho2020denoising}] Denoising Diffusion Probabilistic Model (DDPM) is a method for generating images by reversing a diffusion process. The model progressively denoises random noise to recover data samples. DDPM has shown impressive results in generating high-quality synthetic images and serves as a foundation for various diffusion-based generative models.
    \item[LoRA~\cite{hu2021lora}] LoRA introduces a lightweight fine-tuning method for large models by learning low-rank adaptation. It's often used in diffusion models and provides an efficient way to apply fine-tuning on custom datasets with minimal computation and memory cost. 
    \item[Textual Inversion~\cite{gal2022image}] Textual Inversion (TI) is a method for fine-tuning text-to-image models where specific text tokens are learned to represent new visual concepts. It allows the model to generate images of novel concepts without updating the underlying architecture.
    \item[DCDM~\cite{zhao2023dataset}] Data condensation with distribution matching (DCDM) utilizes distribution matching to condense large datasets into smaller and more efficient synthetic datasets while maintaining a similar statistical distribution as the real data.
    \item[DP-Sinkhorn~\cite{cao2021don}] This method applies DP to generators using Sinkhorn divergence to improve feature matching between real and synthetic datasets. It ensures that synthetic data closely mimics the statistical properties of real data while maintaining DP guarantees.
    \item[DP-LDMs~\cite{lyu2023differentially}] DP-LDM is a DP variant of latent diffusion models that enhances privacy by fine-tuning the attention modules rather than the entire network. Updating minimal parameters reduces noise while maintaining utility in image generation tasks.
    \item[SMP-LoRA~\cite{luo2024privacy}] This method combines LoRA with adversarial training to balance privacy and utility by protecting diffusion models against MIAs. It directly aims to improve the utility-privacy tradeoff as measured by image synthesis loss and MIA attack loss. 
    
\end{description}

\subsection{Hyperparameter Tuning Details and Selection}\label{sec:appendix-hparams}
We perform hyperparameter tuning for each generative method to explore the balance between utility (\emph{e.g.}, classification accuracy) and privacy (\emph{e.g.}, Attack Success Rate). Table~\ref{tab:hyperparameters_tuning} summarizes the specific hyperparameters and search values for each method. In particular, we highlight:

\begin{description}
    \item[GAN (FreezeD)] We freeze the last \emph{N} layers of the discriminator (``Last-1'' to ``Last-5'') to reduce overfitting on small datasets to balance quality vs. privacy.
    \item[VAE] The KL-divergence weight (\(\beta\)) spans \{0.001--0.0002\}; a higher \(\beta\) enforces stronger regularization at the cost of blurrier images, whereas a lower \(\beta\) can overfit but yield sharper samples (with higher privacy risk).
    \item[DDPM] The noise injection strength (0.5--0.8) controls how aggressively partial diffusion is applied. Higher noise improves privacy but can degrade fidelity.
    \item[LoRA] The rank (\(r\in\{16\dots256\}\)) determines how much capacity is allocated for fine-tuning diffusion models. Lower ranks can help privacy; higher ranks may boost accuracy but risk more leakage.
    \item[TI] We vary the classifier-free guidance scale (\{1.0, 2.5, 4.0, 5.5, 7.0\}) to adjust the tradeoff between image realism and adherence to text prompts.
    \item[DCDM] The ratio of real vs.\ synthetic images (\{50\%\dots90\%\}) balances dataset condensation. Greater synthetic proportions can intensify privacy concerns yet offer flexible generation.
    \item[DP-Sinkhorn, DP-LDMs] We vary the privacy budget \(\epsilon \in \{10,50,100,250,500\}\) to explore tradeoffs between privacy guarantees and visual fidelity.
    \item[SMP-LoRA] An adversarial weight \(\lambda\in\{1.0\dots0.01\}\) further emphasizes privacy objectives. Larger \(\lambda\) lowers Attack Success Rate but can reduce classification accuracy, while smaller \(\lambda\) favors fidelity at potential privacy cost.
\end{description}

\noindent
After testing all combinations via a two-fold cross-validation, we select the configuration that yields the highest utility with acceptable privacy risks. The final chosen hyperparameters for each method are used in Figure~\ref{fig:results}.

\begin{table*}[ht]
\centering
\scriptsize
\caption{Results of the impact of hyperparameters on the utility-privacy tradeoff for the CelebA dataset. A1: Attack 1, A2: Attack 2.}
\begin{tabular}{l|c|c|c|c|c}
\hline
\textbf{Method} & \textbf{A1-AUC} & \textbf{A1-TPR@1\%FPR} & \textbf{A2-AUC} & \textbf{A2-TPR@1\%FPR} & \textbf{F1} \\ \hline
DP-SGD ($\epsilon=10$) & - & - & $0.522 \pm 0.009$ & $0.012 \pm 0.003$ & $0.798 \pm 0.005$ \\ \hline
DP-SGD ($\epsilon=50$) & - & - & $0.542 \pm 0.005$ & $0.017 \pm 0.007$ & $0.834 \pm 0.011$ \\ \hline
DP-SGD ($\epsilon=100$) & - & - & $0.566 \pm 0.002$ & $0.015 \pm 0.007$ & $0.859 \pm 0.011$ \\ \hline
DP-SGD ($\epsilon=1000$) & - & - & $0.566 \pm 0.009$ & $0.019 \pm 0.005$ & $0.882 \pm 0.007$ \\ \hline
DP-SGD ($\epsilon=\infty$) & $1.000 \pm 0.000$ & $1.000 \pm 0.000$ & $0.667 \pm 0.037$ & $0.028 \pm 0.007$ & $0.886 \pm 0.005$ \\ \hline
GAN (Last 1) & $0.729 \pm 0.007$ & $0.034 \pm 0.007$ & $0.626 \pm 0.015$ & $0.014 \pm 0.001$ & $0.816 \pm 0.007$ \\ \hline
GAN (Last 2) & $0.727 \pm 0.004$ & $0.040 \pm 0.003$ & $0.637 \pm 0.010$ & $0.078 \pm 0.067$ & $0.821 \pm 0.001$ \\ \hline
GAN (Last 3) & $0.742 \pm 0.016$ & $0.045 \pm 0.006$ & $0.634 \pm 0.015$ & $0.113 \pm 0.004$ & $0.830 \pm 0.007$ \\ \hline
GAN (Last 4) & $0.755 \pm 0.012$ & $0.045 \pm 0.002$ & $0.649 \pm 0.020$ & $0.085 \pm 0.072$ & $0.822 \pm 0.005$ \\ \hline
GAN (Last 5) & $0.759 \pm 0.006$ & $0.034 \pm 0.001$ & $0.638 \pm 0.020$ & $0.042 \pm 0.030$ & $0.826 \pm 0.003$ \\ \hline
VAE ($w=0.001$) & $0.781 \pm 0.001$ & $0.142 \pm 0.014$ & $0.673 \pm 0.018$ & $0.014 \pm 0.002$ & $0.817 \pm 0.007$ \\ \hline
VAE ($w=0.0008$) & $0.797 \pm 0.004$ & $0.174 \pm 0.001$ & $0.695 \pm 0.017$ & $0.170 \pm 0.027$ & $0.821 \pm 0.013$ \\ \hline
VAE ($w=0.0006$) & $0.768 \pm 0.142$ & $0.110 \pm 0.083$ & $0.707 \pm 0.003$ & $0.015 \pm 0.003$ & $0.804 \pm 0.033$ \\ \hline
VAE ($w=0.0004$) & $0.823 \pm 0.007$ & $0.232 \pm 0.008$ & $0.726 \pm 0.024$ & $0.021 \pm 0.004$ & $0.819 \pm 0.009$ \\ \hline
VAE ($w=0.0002$) & $0.846 \pm 0.004$ & $0.267 \pm 0.030$ & $0.739 \pm 0.022$ & $0.023 \pm 0.010$ & $0.814 \pm 0.014$ \\ \hline
DDPM ($\alpha=0.6$) & $0.839 \pm 0.062$ & $0.252 \pm 0.121$ & $0.659 \pm 0.023$ & $0.176 \pm 0.157$ & $0.843 \pm 0.023$ \\ \hline
DDPM ($\alpha=0.7$) & $0.794 \pm 0.105$ & $0.205 \pm 0.133$ & $0.642 \pm 0.001$ & $0.043 \pm 0.032$ & $0.840 \pm 0.016$ \\ \hline
DDPM ($\alpha=0.8$) & $0.743 \pm 0.126$ & $0.168 \pm 0.129$ & $0.627 \pm 0.001$ & $0.170 \pm 0.136$ & $0.842 \pm 0.019$ \\ \hline
DDPM ($\alpha=0.9$) & $0.672 \pm 0.132$ & $0.119 \pm 0.085$ & $0.630 \pm 0.012$ & $0.065 \pm 0.046$ & $0.824 \pm 0.010$ \\ \hline
DDPM ($\alpha=1.0$) & $0.646 \pm 0.096$ & $0.089 \pm 0.054$ & $0.621 \pm 0.018$ & $0.016 \pm 0.002$ & $0.800 \pm 0.019$ \\ \hline
LoRA ($r=16$) & $0.741 \pm 0.008$ & $0.073 \pm 0.000$ & $0.689 \pm 0.003$ & $0.117 \pm 0.109$ & $0.857 \pm 0.014$ \\ \hline
LoRA ($r=32$) & $0.750 \pm 0.020$ & $0.079 \pm 0.007$ & $0.685 \pm 0.002$ & $0.068 \pm 0.053$ & $0.853 \pm 0.020$ \\ \hline
LoRA ($r=64$) & $0.771 \pm 0.009$ & $0.073 \pm 0.004$ & $0.682 \pm 0.005$ & $0.012 \pm 0.002$ & $0.862 \pm 0.018$ \\ \hline
LoRA ($r=128$) & $0.802 \pm 0.013$ & $0.083 \pm 0.001$ & $0.693 \pm 0.009$ & $0.016 \pm 0.000$ & $0.870 \pm 0.032$ \\ \hline
LoRA ($r=256$) & $0.823 \pm 0.027$ & $0.087 \pm 0.023$ & $0.710 \pm 0.017$ & $0.012 \pm 0.003$ & $0.872 \pm 0.027$ \\ \hline
TI ($w=1.0$) & $0.721 \pm 0.007$ & $0.133 \pm 0.023$ & $0.655 \pm 0.006$ & $0.073 \pm 0.004$ & $0.844 \pm 0.006$ \\ \hline
TI ($w=2.5$) & $0.705 \pm 0.006$ & $0.118 \pm 0.027$ & $0.661 \pm 0.007$ & $0.018 \pm 0.003$ & $0.843 \pm 0.008$ \\ \hline
TI ($w=4.0$) & $0.714 \pm 0.009$ & $0.134 \pm 0.005$ & $0.650 \pm 0.014$ & $0.030 \pm 0.005$ & $0.840 \pm 0.013$ \\ \hline
TI ($w=5.5$) & $0.709 \pm 0.006$ & $0.120 \pm 0.006$ & $0.647 \pm 0.004$ & $0.103 \pm 0.090$ & $0.841 \pm 0.011$ \\ \hline
TI ($w=7.0$) & $0.719 \pm 0.002$ & $0.134 \pm 0.010$ & $0.676 \pm 0.007$ & $0.115 \pm 0.093$ & $0.845 \pm 0.012$ \\ \hline
DCDM ($r=0.5$) & $0.740 \pm 0.021$ & $0.429 \pm 0.030$ & $0.646 \pm 0.001$ & $0.120 \pm 0.098$ & $0.858 \pm 0.005$ \\ \hline
DCDM ($r=0.6$) & $0.801 \pm 0.005$ & $0.497 \pm 0.029$ & $0.652 \pm 0.004$ & $0.025 \pm 0.002$ & $0.882 \pm 0.000$ \\ \hline
DCDM ($r=0.7$) & $0.847 \pm 0.001$ & $0.643 \pm 0.002$ & $0.664 \pm 0.009$ & $0.020 \pm 0.002$ & $0.883 \pm 0.005$ \\ \hline
DCDM ($r=0.8$) & $0.909 \pm 0.006$ & $0.750 \pm 0.035$ & $0.695 \pm 0.023$ & $0.091 \pm 0.073$ & $0.881 \pm 0.001$ \\ \hline
DCDM ($r=0.9$) & $0.950 \pm 0.001$ & $0.865 \pm 0.015$ & $0.736 \pm 0.004$ & $0.022 \pm 0.001$ & $0.879 \pm 0.006$ \\ \hline
DP-Sinkhorn ($\epsilon=10$) & $0.536 \pm 0.001$ & $0.034 \pm 0.006$ & $0.536 \pm 0.003$ & $0.105 \pm 0.087$ & $0.789 \pm 0.038$ \\ \hline
DP-Sinkhorn ($\epsilon=50$) & $0.547 \pm 0.003$ & $0.032 \pm 0.012$ & $0.547 \pm 0.004$ & $0.174 \pm 0.025$ & $0.797 \pm 0.041$ \\ \hline
DP-Sinkhorn ($\epsilon=100$) & $0.545 \pm 0.001$ & $0.031 \pm 0.006$ & $0.545 \pm 0.016$ & $0.229 \pm 0.017$ & $0.799 \pm 0.011$ \\ \hline
DP-Sinkhorn ($\epsilon=250$) & $0.559 \pm 0.001$ & $0.028 \pm 0.002$ & $0.559 \pm 0.002$ & $0.022 \pm 0.008$ & $0.816 \pm 0.007$ \\ \hline
DP-Sinkhorn ($\epsilon=500$) & $0.562 \pm 0.003$ & $0.029 \pm 0.001$ & $0.562 \pm 0.004$ & $0.111 \pm 0.098$ & $0.823 \pm 0.013$ \\ \hline
DP-LDMs ($\epsilon=10$) & $0.664 \pm 0.008$ & $0.070 \pm 0.003$ & $0.664 \pm 0.004$ & $0.150 \pm 0.140$ & $0.816 \pm 0.016$ \\ \hline
DP-LDMs ($\epsilon=50$) & $0.669 \pm 0.014$ & $0.075 \pm 0.011$ & $0.676 \pm 0.010$ & $0.013 \pm 0.000$ & $0.822 \pm 0.013$ \\ \hline
DP-LDMs ($\epsilon=100$) & $0.675 \pm 0.013$ & $0.067 \pm 0.008$ & $0.664 \pm 0.010$ & $0.036 \pm 0.025$ & $0.824 \pm 0.008$ \\ \hline
DP-LDMs ($\epsilon=250$) & $0.672 \pm 0.020$ & $0.071 \pm 0.001$ & $0.664 \pm 0.001$ & $0.053 \pm 0.040$ & $0.831 \pm 0.008$ \\ \hline
DP-LDMs ($\epsilon=500$) & $0.673 \pm 0.023$ & $0.088 \pm 0.013$ & $0.679 \pm 0.009$ & $0.016 \pm 0.004$ & $0.837 \pm 0.011$ \\ \hline
SMP-LoRA ($\lambda=1.0$) & $0.680 \pm 0.015$ & $0.087 \pm 0.017$ & $0.682 \pm 0.028$ & $0.018 \pm 0.002$ & $0.837 \pm 0.001$ \\ \hline
SMP-LoRA ($\lambda=0.5$) & $0.674 \pm 0.024$ & $0.106 \pm 0.047$ & $0.699 \pm 0.016$ & $0.016 \pm 0.000$ & $0.833 \pm 0.005$ \\ \hline
SMP-LoRA ($\lambda=0.1$) & $0.711 \pm 0.064$ & $0.129 \pm 0.065$ & $0.675 \pm 0.012$ & $0.013 \pm 0.001$ & $0.834 \pm 0.009$ \\ \hline
SMP-LoRA ($\lambda=0.05$) & $0.728 \pm 0.100$ & $0.165 \pm 0.113$ & $0.688 \pm 0.007$ & $0.017 \pm 0.000$ & $0.837 \pm 0.006$ \\ \hline
SMP-LoRA ($\lambda=0.01$) & $0.762 \pm 0.112$ & $0.243 \pm 0.143$ & $0.695 \pm 0.007$ & $0.017 \pm 0.003$ & $0.851 \pm 0.001$ \\ \hline
\end{tabular}
\label{tab:celeba}
\end{table*}

\begin{table*}[ht]
\centering
\scriptsize
\caption{Results of the impact of hyperparameters on the utility-privacy tradeoff for the Fitzpatrick17k dataset. A1: Attack 1, A2: Attack 2.}
\begin{tabular}{l|c|c|c|c|c}
\hline
\textbf{Method} & \textbf{A1-AUC} & \textbf{A1-TPR@1\%FPR} & \textbf{A2-AUC} & \textbf{A2-TPR@1\%FPR} & \textbf{Accuracy} \\ \hline
DP-SGD ($\epsilon=10$) & - & - & $0.537 \pm 0.011$ & $0.056 \pm 0.010$ & $0.552 \pm 0.011$ \\ \hline
DP-SGD ($\epsilon=50$) & - & - & $0.567 \pm 0.028$ & $0.029 \pm 0.011$ & $0.651 \pm 0.010$ \\ \hline
DP-SGD ($\epsilon=100$) & - & - & $0.585 \pm 0.023$ & $0.024 \pm 0.012$ & $0.671 \pm 0.010$ \\ \hline
DP-SGD ($\epsilon=5000$) & - & - & $0.707 \pm 0.022$ & $0.036 \pm 0.022$ & $0.707 \pm 0.014$ \\ \hline
DP-SGD ($\epsilon=7000$) & - & - & $0.717 \pm 0.026$ & $0.041 \pm 0.014$ & $0.720 \pm 0.020$ \\ \hline
DP-SGD ($\epsilon=\infty$) & $1.000 \pm 0.000$ & $1.000 \pm 0.000$ & $0.739 \pm 0.024$ & $0.056 \pm 0.018$ & $0.734 \pm 0.008$ \\ \hline
GAN (Last 1) & $0.729 \pm 0.006$ & $0.030 \pm 0.009$ & $0.626 \pm 0.009$ & $0.034 \pm 0.009$ & $0.676 \pm 0.002$ \\ \hline
GAN (Last 2) & $0.727 \pm 0.006$ & $0.032 \pm 0.007$ & $0.637 \pm 0.001$ & $0.030 \pm 0.001$ & $0.675 \pm 0.003$ \\ \hline
GAN (Last 3) & $0.742 \pm 0.004$ & $0.028 \pm 0.007$ & $0.634 \pm 0.009$ & $0.020 \pm 0.002$ & $0.666 \pm 0.001$ \\ \hline
GAN (Last 4) & $0.755 \pm 0.004$ & $0.032 \pm 0.012$ & $0.649 \pm 0.010$ & $0.024 \pm 0.011$ & $0.685 \pm 0.011$ \\ \hline
GAN (Last 5) & $0.759 \pm 0.001$ & $0.028 \pm 0.002$ & $0.638 \pm 0.007$ & $0.036 \pm 0.014$ & $0.663 \pm 0.004$ \\ \hline
VAE ($w=0.001$) & $0.781 \pm 0.015$ & $0.052 \pm 0.004$ & $0.673 \pm 0.001$ & $0.062 \pm 0.018$ & $0.690 \pm 0.010$ \\ \hline
VAE ($w=0.0008$) & $0.797 \pm 0.010$ & $0.055 \pm 0.003$ & $0.695 \pm 0.001$ & $0.063 \pm 0.010$ & $0.685 \pm 0.015$ \\ \hline
VAE ($w=0.0006$) & $0.768 \pm 0.009$ & $0.071 \pm 0.006$ & $0.707 \pm 0.012$ & $0.070 \pm 0.025$ & $0.702 \pm 0.003$ \\ \hline
VAE ($w=0.0004$) & $0.823 \pm 0.014$ & $0.074 \pm 0.011$ & $0.726 \pm 0.000$ & $0.065 \pm 0.009$ & $0.692 \pm 0.000$ \\ \hline
VAE ($w=0.0002$) & $0.846 \pm 0.012$ & $0.101 \pm 0.005$ & $0.739 \pm 0.003$ & $0.058 \pm 0.003$ & $0.706 \pm 0.004$ \\ \hline
DDPM ($\alpha=0.6$) & $0.839 \pm 0.058$ & $0.065 \pm 0.035$ & $0.659 \pm 0.018$ & $0.076 \pm 0.018$ & $0.700 \pm 0.027$ \\ \hline
DDPM ($\alpha=0.7$) & $0.794 \pm 0.053$ & $0.055 \pm 0.029$ & $0.642 \pm 0.020$ & $0.054 \pm 0.014$ & $0.681 \pm 0.024$ \\ \hline
DDPM ($\alpha=0.8$) & $0.743 \pm 0.030$ & $0.044 \pm 0.014$ & $0.627 \pm 0.022$ & $0.084 \pm 0.006$ & $0.680 \pm 0.030$ \\ \hline
DDPM ($\alpha=0.9$) & $0.672 \pm 0.037$ & $0.059 \pm 0.026$ & $0.630 \pm 0.032$ & $0.065 \pm 0.024$ & $0.671 \pm 0.025$ \\ \hline
DDPM ($\alpha=1.0$) & $0.646 \pm 0.033$ & $0.043 \pm 0.022$ & $0.621 \pm 0.015$ & $0.080 \pm 0.021$ & $0.689 \pm 0.030$ \\ \hline
LoRA ($r=16$) & $0.694 \pm 0.003$ & $0.086 \pm 0.008$ & $0.689 \pm 0.024$ & $0.054 \pm 0.003$ & $0.720 \pm 0.020$ \\ \hline
LoRA ($r=32$) & $0.709 \pm 0.007$ & $0.078 \pm 0.009$ & $0.685 \pm 0.009$ & $0.074 \pm 0.019$ & $0.733 \pm 0.014$ \\ \hline
LoRA ($r=64$) & $0.700 \pm 0.027$ & $0.098 \pm 0.021$ & $0.682 \pm 0.022$ & $0.044 \pm 0.010$ & $0.740 \pm 0.006$ \\ \hline
LoRA ($r=128$) & $0.721 \pm 0.008$ & $0.078 \pm 0.001$ & $0.693 \pm 0.023$ & $0.048 \pm 0.004$ & $0.746 \pm 0.009$ \\ \hline
LoRA ($r=256$) & $0.738 \pm 0.008$ & $0.087 \pm 0.003$ & $0.710 \pm 0.001$ & $0.075 \pm 0.004$ & $0.744 \pm 0.014$ \\ \hline
TI ($w=1.0$) & $0.681 \pm 0.025$ & $0.049 \pm 0.012$ & $0.655 \pm 0.047$ & $0.041 \pm 0.007$ & $0.700 \pm 0.038$ \\ \hline
TI ($w=2.5$) & $0.665 \pm 0.020$ & $0.048 \pm 0.001$ & $0.661 \pm 0.056$ & $0.078 \pm 0.044$ & $0.713 \pm 0.054$ \\ \hline
TI ($w=4.0$) & $0.674 \pm 0.023$ & $0.053 \pm 0.015$ & $0.650 \pm 0.015$ & $0.035 \pm 0.007$ & $0.708 \pm 0.042$ \\ \hline
TI ($w=5.5$) & $0.669 \pm 0.021$ & $0.046 \pm 0.002$ & $0.647 \pm 0.062$ & $0.034 \pm 0.023$ & $0.715 \pm 0.047$ \\ \hline
TI ($w=7.0$) & $0.679 \pm 0.020$ & $0.044 \pm 0.007$ & $0.676 \pm 0.049$ & $0.049 \pm 0.003$ & $0.718 \pm 0.041$ \\ \hline
DCDM ($r=0.5$) & $0.733 \pm 0.001$ & $0.248 \pm 0.030$ & $0.646 \pm 0.019$ & $0.051 \pm 0.004$ & $0.712 \pm 0.005$ \\ \hline
DCDM ($r=0.6$) & $0.767 \pm 0.014$ & $0.335 \pm 0.030$ & $0.652 \pm 0.013$ & $0.055 \pm 0.007$ & $0.714 \pm 0.013$ \\ \hline
DCDM ($r=0.7$) & $0.834 \pm 0.011$ & $0.425 \pm 0.002$ & $0.664 \pm 0.024$ & $0.051 \pm 0.030$ & $0.713 \pm 0.008$ \\ \hline
DCDM ($r=0.8$) & $0.889 \pm 0.007$ & $0.533 \pm 0.028$ & $0.695 \pm 0.019$ & $0.076 \pm 0.008$ & $0.735 \pm 0.008$ \\ \hline
DCDM ($r=0.9$) & $0.946 \pm 0.002$ & $0.664 \pm 0.028$ & $0.736 \pm 0.002$ & $0.045 \pm 0.009$ & $0.731 \pm 0.011$ \\ \hline
DP-Sinkhorn ($\epsilon=10$) & $0.536 \pm 0.004$ & $0.037 \pm 0.012$ & $0.536 \pm 0.012$ & $0.021 \pm 0.002$ & $0.528 \pm 0.018$ \\ \hline
DP-Sinkhorn ($\epsilon=50$) & $0.547 \pm 0.001$ & $0.030 \pm 0.003$ & $0.547 \pm 0.003$ & $0.012 \pm 0.003$ & $0.536 \pm 0.021$ \\ \hline
DP-Sinkhorn ($\epsilon=100$) & $0.545 \pm 0.009$ & $0.029 \pm 0.008$ & $0.545 \pm 0.008$ & $0.069 \pm 0.031$ & $0.566 \pm 0.002$ \\ \hline
DP-Sinkhorn ($\epsilon=250$) & $0.559 \pm 0.016$ & $0.023 \pm 0.005$ & $0.559 \pm 0.022$ & $0.063 \pm 0.039$ & $0.571 \pm 0.015$ \\ \hline
DP-Sinkhorn ($\epsilon=500$) & $0.562 \pm 0.008$ & $0.036 \pm 0.012$ & $0.562 \pm 0.017$ & $0.017 \pm 0.001$ & $0.567 \pm 0.011$ \\ \hline
DP-LDMs ($\epsilon=10$) & $0.664 \pm 0.007$ & $0.078 \pm 0.009$ & $0.664 \pm 0.019$ & $0.053 \pm 0.006$ & $0.711 \pm 0.007$ \\ \hline
DP-LDMs ($\epsilon=50$) & $0.662 \pm 0.021$ & $0.095 \pm 0.018$ & $0.676 \pm 0.007$ & $0.082 \pm 0.010$ & $0.719 \pm 0.008$ \\ \hline
DP-LDMs ($\epsilon=100$) & $0.659 \pm 0.013$ & $0.084 \pm 0.007$ & $0.664 \pm 0.001$ & $0.080 \pm 0.005$ & $0.705 \pm 0.012$ \\ \hline
DP-LDMs ($\epsilon=250$) & $0.662 \pm 0.017$ & $0.090 \pm 0.002$ & $0.664 \pm 0.015$ & $0.088 \pm 0.020$ & $0.716 \pm 0.008$ \\ \hline
DP-LDMs ($\epsilon=500$) & $0.663 \pm 0.012$ & $0.086 \pm 0.004$ & $0.679 \pm 0.018$ & $0.076 \pm 0.027$ & $0.720 \pm 0.019$ \\ \hline
SMP-LoRA ($\lambda=1.0$) & $0.620 \pm 0.015$ & $0.057 \pm 0.008$ & $0.682 \pm 0.042$ & $0.052 \pm 0.014$ & $0.701 \pm 0.014$ \\ \hline
SMP-LoRA ($\lambda=0.5$) & $0.638 \pm 0.017$ & $0.069 \pm 0.010$ & $0.699 \pm 0.022$ & $0.077 \pm 0.009$ & $0.737 \pm 0.007$ \\ \hline
SMP-LoRA ($\lambda=0.1$) & $0.654 \pm 0.010$ & $0.090 \pm 0.007$ & $0.675 \pm 0.035$ & $0.059 \pm 0.003$ & $0.742 \pm 0.003$ \\ \hline
SMP-LoRA ($\lambda=0.05$) & $0.678 \pm 0.004$ & $0.122 \pm 0.006$ & $0.688 \pm 0.005$ & $0.073 \pm 0.014$ & $0.734 \pm 0.002$ \\ \hline
SMP-LoRA ($\lambda=0.01$) & $0.738 \pm 0.002$ & $0.218 \pm 0.013$ & $0.695 \pm 0.005$ & $0.059 \pm 0.006$ & $0.734 \pm 0.011$ \\ \hline
\end{tabular}
\label{tab:fitzpatrick}
\end{table*}

\begin{table*}[ht]
\centering
\scriptsize
\caption{Results of the impact of hyperparameters on the utility-privacy tradeoff for the CheXpert dataset. A1: Attack 1, A2: Attack 2.}
\begin{tabular}{l|c|c|c|c|c}
\hline
\textbf{Method} & \textbf{A1-AUC} & \textbf{A1-TPR@1\%FPR} & \textbf{A2-AUC} & \textbf{A2-TPR@1\%FPR} & \textbf{AUC} \\ \hline
DP-SGD ($\epsilon=10$) & - & - & $0.523 \pm 0.021$ & $0.020 \pm 0.011$ & $0.529 \pm 0.015$ \\ \hline
DP-SGD ($\epsilon=50$) & - & - & $0.542 \pm 0.024$ & $0.030 \pm 0.015$ & $0.551 \pm 0.017$ \\ \hline
DP-SGD ($\epsilon=100$) & - & - & $0.567 \pm 0.014$ & $0.036 \pm 0.015$ & $0.547 \pm 0.011$ \\ \hline
DP-SGD ($\epsilon=1000$) & - & - & $0.643 \pm 0.018$ & $0.060 \pm 0.013$ & $0.605 \pm 0.030$ \\ \hline
DP-SGD ($\epsilon=\infty$) & $1.000 \pm 0.000$ & $1.000 \pm 0.000$ & $0.989 \pm 0.016$ & $0.766 \pm 0.320$ & $0.692 \pm 0.018$ \\ \hline
GAN (Last 1) & $0.511 \pm 0.003$ & $0.024 \pm 0.007$ & $0.817 \pm 0.015$ & $0.166 \pm 0.009$ & $0.670 \pm 0.014$ \\ \hline
GAN (Last 2) & $0.530 \pm 0.005$ & $0.030 \pm 0.007$ & $0.851 \pm 0.011$ & $0.350 \pm 0.012$ & $0.673 \pm 0.017$ \\ \hline
GAN (Last 3) & $0.563 \pm 0.008$ & $0.037 \pm 0.009$ & $0.870 \pm 0.014$ & $0.166 \pm 0.066$ & $0.668 \pm 0.011$ \\ \hline
GAN (Last 4) & $0.572 \pm 0.014$ & $0.053 \pm 0.010$ & $0.864 \pm 0.003$ & $0.353 \pm 0.016$ & $0.673 \pm 0.007$ \\ \hline
GAN (Last 5) & $0.558 \pm 0.023$ & $0.038 \pm 0.001$ & $0.838 \pm 0.018$ & $0.203 \pm 0.072$ & $0.677 \pm 0.010$ \\ \hline
VAE ($w=0.001$) & $0.536 \pm 0.016$ & $0.029 \pm 0.000$ & $0.967 \pm 0.024$ & $0.506 \pm 0.256$ & $0.676 \pm 0.011$ \\ \hline
VAE ($w=0.0008$) & $0.559 \pm 0.012$ & $0.036 \pm 0.008$ & $0.986 \pm 0.009$ & $0.688 \pm 0.094$ & $0.679 \pm 0.009$ \\ \hline
VAE ($w=0.0006$) & $0.550 \pm 0.003$ & $0.032 \pm 0.000$ & $0.982 \pm 0.023$ & $0.597 \pm 0.350$ & $0.683 \pm 0.017$ \\ \hline
VAE ($w=0.0004$) & $0.551 \pm 0.009$ & $0.032 \pm 0.002$ & $0.972 \pm 0.013$ & $0.513 \pm 0.284$ & $0.680 \pm 0.011$ \\ \hline
VAE ($w=0.0002$) & $0.538 \pm 0.010$ & $0.032 \pm 0.005$ & $0.980 \pm 0.003$ & $0.606 \pm 0.062$ & $0.676 \pm 0.015$ \\ \hline
DDPM ($\alpha=0.6$) & $0.537 \pm 0.017$ & $0.029 \pm 0.005$ & $0.822 \pm 0.156$ & $0.406 \pm 0.250$ & $0.676 \pm 0.016$ \\ \hline
DDPM ($\alpha=0.7$) & $0.517 \pm 0.001$ & $0.026 \pm 0.004$ & $0.739 \pm 0.000$ & $0.088 \pm 0.013$ & $0.673 \pm 0.010$ \\ \hline
DDPM ($\alpha=0.8$) & $0.513 \pm 0.008$ & $0.030 \pm 0.009$ & $0.753 \pm 0.120$ & $0.181 \pm 0.125$ & $0.666 \pm 0.013$ \\ \hline
DDPM ($\alpha=0.9$) & $0.539 \pm 0.001$ & $0.031 \pm 0.002$ & $0.739 \pm 0.125$ & $0.119 \pm 0.081$ & $0.662 \pm 0.015$ \\ \hline
DDPM ($\alpha=1.0$) & $0.518 \pm 0.008$ & $0.029 \pm 0.006$ & $0.797 \pm 0.146$ & $0.456 \pm 0.325$ & $0.671 \pm 0.003$ \\ \hline
LoRA ($r=16$) & $0.618 \pm 0.020$ & $0.040 \pm 0.010$ & $0.962 \pm 0.008$ & $0.575 \pm 0.119$ & $0.677 \pm 0.007$ \\ \hline
LoRA ($r=32$) & $0.687 \pm 0.011$ & $0.066 \pm 0.007$ & $0.981 \pm 0.004$ & $0.703 \pm 0.041$ & $0.678 \pm 0.008$ \\ \hline
LoRA ($r=64$) & $0.716 \pm 0.057$ & $0.102 \pm 0.054$ & $0.974 \pm 0.005$ & $0.644 \pm 0.062$ & $0.685 \pm 0.010$ \\ \hline
LoRA ($r=128$) & $0.788 \pm 0.000$ & $0.153 \pm 0.056$ & $0.944 \pm 0.016$ & $0.406 \pm 0.088$ & $0.683 \pm 0.008$ \\ \hline
LoRA ($r=256$) & $0.859 \pm 0.017$ & $0.240 \pm 0.010$ & $0.970 \pm 0.008$ & $0.512 \pm 0.094$ & $0.690 \pm 0.007$ \\ \hline
TI ($w=1.0$) & $0.626 \pm 0.001$ & $0.061 \pm 0.001$ & $0.864 \pm 0.031$ & $0.134 \pm 0.041$ & $0.692 \pm 0.023$ \\ \hline
TI ($w=2.5$) & $0.623 \pm 0.008$ & $0.064 \pm 0.010$ & $0.876 \pm 0.038$ & $0.134 \pm 0.009$ & $0.695 \pm 0.020$ \\ \hline
TI ($w=4.0$) & $0.628 \pm 0.002$ & $0.051 \pm 0.009$ & $0.893 \pm 0.031$ & $0.303 \pm 0.097$ & $0.693 \pm 0.023$ \\ \hline
TI ($w=5.5$) & $0.627 \pm 0.012$ & $0.057 \pm 0.001$ & $0.849 \pm 0.034$ & $0.125 \pm 0.044$ & $0.696 \pm 0.014$ \\ \hline
TI ($w=7.0$) & $0.610 \pm 0.011$ & $0.058 \pm 0.011$ & $0.897 \pm 0.039$ & $0.268 \pm 0.125$ & $0.690 \pm 0.024$ \\ \hline
DCDM ($r=0.5$) & $0.641 \pm 0.003$ & $0.091 \pm 0.002$ & $0.825 \pm 0.006$ & $0.431 \pm 0.019$ & $0.679 \pm 0.013$ \\ \hline
DCDM ($r=0.6$) & $0.737 \pm 0.010$ & $0.208 \pm 0.009$ & $0.877 \pm 0.008$ & $0.559 \pm 0.028$ & $0.680 \pm 0.003$ \\ \hline
DCDM ($r=0.7$) & $0.798 \pm 0.010$ & $0.353 \pm 0.036$ & $0.917 \pm 0.019$ & $0.572 \pm 0.122$ & $0.682 \pm 0.013$ \\ \hline
DCDM ($r=0.8$) & $0.900 \pm 0.002$ & $0.562 \pm 0.079$ & $0.898 \pm 0.031$ & $0.616 \pm 0.197$ & $0.686 \pm 0.017$ \\ \hline
DCDM ($r=0.9$) & $0.938 \pm 0.006$ & $0.710 \pm 0.064$ & $0.955 \pm 0.026$ & $0.831 \pm 0.056$ & $0.679 \pm 0.010$ \\ \hline
DP-Sinkhorn ($\epsilon=10$) & $0.522 \pm 0.005$ & $0.027 \pm 0.003$ & $0.649 \pm 0.048$ & $0.022 \pm 0.022$ & $0.649 \pm 0.019$ \\ \hline
DP-Sinkhorn ($\epsilon=50$) & $0.523 \pm 0.004$ & $0.024 \pm 0.002$ & $0.649 \pm 0.025$ & $0.050 \pm 0.019$ & $0.649 \pm 0.012$ \\ \hline
DP-Sinkhorn ($\epsilon=100$) & $0.516 \pm 0.004$ & $0.026 \pm 0.005$ & $0.640 \pm 0.016$ & $0.041 \pm 0.022$ & $0.640 \pm 0.003$ \\ \hline
DP-Sinkhorn ($\epsilon=250$) & $0.525 \pm 0.002$ & $0.026 \pm 0.006$ & $0.625 \pm 0.010$ & $0.047 \pm 0.022$ & $0.625 \pm 0.010$ \\ \hline
DP-Sinkhorn ($\epsilon=500$) & $0.528 \pm 0.012$ & $0.029 \pm 0.001$ & $0.653 \pm 0.018$ & $0.050 \pm 0.006$ & $0.653 \pm 0.003$ \\ \hline
DP-LDMs ($\epsilon=10$)  & $0.578 \pm 0.005$ & $0.035 \pm 0.005$ & $0.972 \pm 0.010$ & $0.509 \pm 0.141$ & $0.677 \pm 0.018$ \\ \hline
DP-LDMs ($\epsilon=50$)  & $0.643 \pm 0.012$ & $0.061 \pm 0.002$ & $0.956 \pm 0.007$ & $0.253 \pm 0.016$ & $0.679 \pm 0.011$ \\ \hline
DP-LDMs ($\epsilon=100$) & $0.709 \pm 0.012$ & $0.063 \pm 0.001$ & $0.976 \pm 0.014$ & $0.550 \pm 0.156$ & $0.678 \pm 0.018$ \\ \hline
DP-LDMs ($\epsilon=250$) & $0.759 \pm 0.000$ & $0.111 \pm 0.007$ & $0.970 \pm 0.014$ & $0.450 \pm 0.112$ & $0.678 \pm 0.016$ \\ \hline
DP-LDMs ($\epsilon=500$) & $0.748 \pm 0.017$ & $0.107 \pm 0.022$ & $0.961 \pm 0.001$ & $0.331 \pm 0.062$ & $0.678 \pm 0.009$ \\ \hline
SMP-LoRA ($\lambda=1.0$) & $0.988 \pm 0.001$ & $0.750 \pm 0.088$ & $0.933 \pm 0.011$ & $0.306 \pm 0.041$ & $0.736 \pm 0.020$ \\ \hline
SMP-LoRA ($\lambda=0.5$) & $0.974 \pm 0.005$ & $0.690 \pm 0.131$ & $0.896 \pm 0.059$ & $0.525 \pm 0.159$ & $0.747 \pm 0.000$ \\ \hline
SMP-LoRA ($\lambda=0.1$) & $0.980 \pm 0.002$ & $0.744 \pm 0.080$ & $0.965 \pm 0.070$ & $0.531 \pm 0.209$ & $0.718 \pm 0.000$ \\ \hline
SMP-LoRA ($\lambda=0.05$) & $0.982 \pm 0.007$ & $0.735 \pm 0.153$ & $0.916 \pm 0.003$ & $0.456 \pm 0.069$ & $0.742 \pm 0.019$ \\ \hline
SMP-LoRA ($\lambda=0.01$) & $0.997 \pm 0.002$ & $0.942 \pm 0.044$ & $0.827 \pm 0.029$ & $0.313 \pm 0.072$ & $0.751 \pm 0.016$ \\ \hline
\end{tabular}
\label{tab:chexpert}
\end{table*}

\paragraph{Computational Cost and Running Times on CheXpert.}
We further analyzed the computational overhead by measuring the total running time for each method on the CheXpert dataset. These times typically encompass generative model training/fine-tuning, synthetic image generation, and downstream classifier training. All experiments were conducted on an NVIDIA A100 GPU. The total running times, converted to GPU hours, are presented in Table~\ref{tab:chexpert_running_times}.

\begin{table*}[htbp] 
\centering
\caption{Total running times for benchmarked methods on the CheXpert dataset, presented in A100 GPU hours.}
\label{tab:chexpert_running_times}
\begin{tabular}{lc}
\hline
Method & Total Running Time (GPU hours) \\
\hline
DP-SGD (Classifier on Real Data) & 0.50 \\
VAE & 1.62 \\
DCDM & 1.30 \\
DP-Sinkhorn & 0.62 \\
DDPM & 3.30 \\
GAN (FreezeD) & 6.80 \\
LoRA & 6.82 \\
DP-LDMs & 5.18 \\
Textual Inversion (TI) & 18.05 \\
SMP-LoRA & 22.38 \\
\hline
\end{tabular}
\end{table*}

\noindent 
These running times highlight the varying computational demands of different approaches. For instance, methods like DP-SGD, DP-Sinkhorn, DCDM, and VAE demonstrate lower computational footprints. In contrast, more complex fine-tuning strategies for large diffusion models, such as Textual Inversion and SMP-LoRA, require substantially more GPU resources. This information is pertinent for practical deployment, especially in scenarios with limited computational budgets.

\paragraph{All Results Across Hyperparameters}
The results of the utility-privacy tradeoff for each method under different hyperparameter configurations are presented in two separate tables: Table~\ref{tab:celeba} for CelebA and Table~\ref{tab:fitzpatrick} for Fitzpatrick17k. Each method explores a unique set of hyperparameters. Adjusting these hyperparameters directly shifts the method’s position along the utility-privacy spectrum. For example, LoRA with rank \(r=16\) achieves moderate utility and lower Attack Success Rate, whereas \(r=256\) yields higher accuracy but also greater privacy leakage. Similarly, SMP-LoRA with a large adversarial weight \(\lambda=1.0\) heavily emphasizes privacy at the expense of classification accuracy. In contrast, \(\lambda=0.01\) focuses on utility to improve accuracy but risks of increased Attack Success Rate. Comparable patterns emerge in other methods such as VAE (where a high KL weight can reduce overfitting but degrade realism) and DDPM (where stronger noise injection can hamper membership inference yet lower fidelity). Overall, these findings highlight that no single hyperparameter works best in every scenario. Model designers must choose an acceptable compromise between privacy and utility. 

\begin{table*}[htbp]
\centering
\caption{Disparate impact between lighter (Scale 2) and darker (Scale 6) skin types on Fitzpatrick17k. The change ($\Delta$) in Scale 2/6 accuracy relative to DP-SGD ($\epsilon=\infty$) are presented.}
\begin{tabular}{l l r r r r r}
\hline
Method & Overall Acc. & Lighter (S2) skin & $\Delta$ S2 vs Base & Darker (S6) skin &  $\Delta$ S6 vs Base \\ 
\hline
DP-SGD ($\epsilon=\infty$) & 0.734 & 0.712 & - & 0.757 & - \\
DP-SGD ($\epsilon=10$) & 0.552 & 0.543 & -0.169 & 0.570 & -0.187 \\
DP-SGD ($\epsilon=100$) & 0.671 & 0.649 & -0.063 & 0.648 & -0.109 \\
DP-SGD ($\epsilon=5000$) & 0.695 & 0.679 & -0.033 & 0.666 & -0.091 \\
DP-SGD ($\epsilon=7000$) & 0.717 & 0.702 & -0.010 & 0.783 & 0.026 \\ \hline
Average & & & -0.069 & & -0.090 \\ \hline
GAN & 0.675 & 0.661 & -0.051 & 0.798 & 0.041 \\
VAE & 0.706 & 0.661 & -0.051 & 0.820 & 0.063 \\
DDPM & 0.681 & 0.648 & -0.064 & 0.698 & -0.059 \\
LoRA & 0.740 & 0.722 & 0.010 & 0.753 & -0.004 \\
TI & 0.715 & 0.685 & -0.027 & 0.774 & 0.017 \\
DCDM & 0.712 & 0.673 & -0.039 & 0.719 & -0.038 \\
DP-Sinkhorn & 0.567 & 0.523 & -0.189 & 0.533 & -0.224 \\
DP-LDMs & 0.716 & 0.692 & -0.020 & 0.711 & -0.046 \\
SMP-LoRA & 0.742 & 0.726 & 0.014 & 0.820 & 0.063 \\ \hline
Average & & & -0.049 & & -0.021 \\ 
\hline
\end{tabular}
\label{tab:fitz_disparate_impact_analysis}
\end{table*}
\end{document}